\documentclass[fleqn,usenatbib]{mnras}

\usepackage{newtxtext,newtxmath}

\usepackage[T1]{fontenc}

\DeclareRobustCommand{\VAN}[3]{#2}
\let\VANthebibliography\thebibliography
\def\thebibliography{\DeclareRobustCommand{\VAN}[3]{##3}\VANthebibliography}

\usepackage{graphicx}	
\usepackage{amsmath}	
\usepackage{bm}         
\usepackage{pdflscape}
\usepackage[dvipsnames]{xcolor}     
\usepackage[normalem]{ulem}

\newcommand{\msun}{\ensuremath{\, \mathrm {\rm M}_{\sun{}}}}
\newcommand{\GW}{\mathrm{GW}}
\newcommand{\BH}{\mathrm{BH}}
\newcommand{\rlx}{\mathrm{rlx}}
\newcommand{\LC}{\mathrm{LC}}

\title[EMRIs and TDEs in nuclear clusters I]{Extreme mass ratio inspirals and tidal disruption events in nuclear clusters. I. Time dependent rates.}

\author[L. Broggi et al.]{
Luca Broggi,$^{1,2}$\thanks{E-mail: l.broggi1@campus.unimib.it}
Elisa Bortolas,$^{1,2}$
Matteo Bonetti,$^{1,2,3}$
Alberto Sesana,$^{1,2,3}$
Massimo Dotti$^{1,2,3}$
\\
$^{1}$Dipartimento di Fisica “G. Occhialini”, Università degli Studi di Milano-Bicocca
Piazza della Scienza 3
20126 Milano, Italy\\
$^{2}$INFN, Sezione di Milano-Bicocca
Piazza della Scienza 3
20126 Milano, Italy\\
$^{3}$INAF - Osservatorio Astronomico di Brera, via Brera 20, 20121 Milano, Italy
}

\date{Accepted XXX. Received YYY; in original form ZZZ}

\pubyear{2022}

\begin{document}
\label{firstpage}
\pagerange{\pageref{firstpage}--\pageref{lastpage}}
\maketitle

\begin{abstract}

In this paper we develop a computationally efficient, two-population, time-dependent Fokker-Plank approach in the two dimensions of energy and angular momentum to study the rates of tidal disruption events (TDEs), extreme mass ratio inspirals (EMRIs) and direct plunges occurring around massive black holes (MBHs) in galactic nuclei. We test our code by exploring a wide range of the astrophysically relevant parameter space, including MBH masses, galaxy central densities and inner density slopes. We find that mass segregation and, more in general, the time dependency of the distribution function regulate the event rate: TDEs always decline with time, whereas EMRIs and plunges  reach a maximum and undergo a subsequent nearly exponential decay. Once suitably normalized, the rates associated to different choices of MBH mass and galaxy density overlap nearly perfectly. Based on this, we provide a simple scaling that allows to reproduce the time-dependent event rates for any MBH mass and underlying galactic nucleus.
 Although our {\it peak} rates are in general agreement with the literature relying on the steady-state (non-time dependent) assumption, those can be sustained on a timescale that strongly depends on the properties of the system. In particular this can be much shorter than a Gyr for relatively light MBHs residing in dense systems. This warns against using steady state models to compute global TDE, EMRI and plunge rates and calls for a more sophisticated, time dependent treatment of the problem.


\end{abstract}




\section{Introduction}

Massive black holes (MBHs) with masses in the range $10^4-10^{10}\msun{}$ are found to reside at the centre of many  galaxies \citep[e.g.][]{1993MNRAS.263..168H,2021ApJ...909...80B} and  
they are often surrounded by very dense, compact stellar systems, whose densities can reach $10^7$ $\msun{}$ pc$^{-3}$, named nuclear star clusters \citep[][]{1997AJ....114.2366C,2020A&ARv..28....4N}. Within such a crowded environment, stars and compact objects are randomly deflected by relaxation processes 
and can be scattered onto very low  angular momentum orbits, closely approaching the central MBHs, and giving rise to violent and exotic phenomena \citep[][]{merritt_dynamics_2013}. 

Depending on the nature of the objects, the outcome of such close interactions can be different. 
If a regular star gets too close to an MBH with mass  $\lesssim10^7\msun{}$, tidal forces rip it apart prompting an episode of efficient (likely super-Eddington, \citealt{2011MNRAS.410..359L}) accretion resulting in a luminous transient flare with a typical light curve \citep[e.g.][]{2009MNRAS.392..332L,2009MNRAS.400.2070S}. A similar fate awaits white dwarfs approaching MBHs of more modest mass, up to $\approx 10^5\msun$ \citep{2008MNRAS.391..718S,2009ApJ...695..404R}. Several of these  tidal disruption events (TDEs, \citealt{1988Natur.333..523R}) have been observed in recent years in the optical, UV and X-ray band \citep[][]{2021SSRv..217...18S,2021ApJ...908....4V}. 

Compact objects, in particular stellar black holes (sBHs), cannot be torn apart by tidal forces, but if they find themselves on an orbit that reaches a  close enough separation to the MBH, they may eventually enter its horizon emitting gravitational waves (GWs) along the way. If the compact object gets deflected on an orbit for which GW emission is significant, but the orbital decay is slow, it may give rise to a detectable, long-lasting GW signal, eventually plunging onto the MBH after many cycles \citep[see][for comprehensive reviews]{2007CQGra..24R.113A,amaro-seoane_relativistic_2018}. If the MBH mass is in the range of $\sim 10^4-10^{7} \msun{}$, the emitted GW signal falls in the mHz frequency window and is anticipated to be one of the primary sources for the forthcoming Laser Interferometer Space Antenna (LISA, \citealt{2017arXiv170200786A,2017PhRvD..95j3012B}). Because of the very unequal mass of the two objects involved in the system, those GW sources are called extreme mass ratio inspirals (EMRIs). Besides producing EMRIs, sBHs can also be deflected onto 'head on collisions' with the central MBHs, directly plunging in the event horizon without experiencing any significant inspiral (and GW emission).
EMRIs will offer an unprecedented way to probe the immediate vicinity of an MBH,  allowing to test 
General Relativity in the strong field regime through the analysis of the emitted GW signals \citep{PhysRevD.75.042003, gair_testing_2013}, unveling the cosmic population of dormant MBHs \citep{2010PhRvD..81j4014G}, and providing a powerful tool to measure the expansion rate of the Universe \citep{2008PhRvD..77d3512M,2021MNRAS.508.4512L}. These  sources can be detected either as  single events \citep[e.g][]{2004PhRvD..69h2005B} or as a cumulative background signal \citep{2004PhRvD..70l2002B,bonetti_gravitational_2020}.
On the other hand, the debris accreted from the many tidally disrupted stars \citep[][]{stone_rates_2016} may dominate the accretion, potentially offering the opportunity for the simultaneous detection of gravitational and electromagnetic emissions \citep{pestoni_generation_2020}.


Although several physical processes -- including tidal separation of binaries \citep{Miller_2005}, perturbations due to a MBH binary \citep{2011ApJ...729...13C,BodeWegg,Naoz2022,2022arXiv220405343M}, capture and migration within AGN accretion disks \citep{2007MNRAS.374..515L,pan_formation_2021} and supernovae explosions \citep{2019MNRAS.485.2125B} -- can significantly contribute to the cosmic TDE, EMRI and plunge rate, the main formation mechanism deals with dynamical relaxation processes within dense galactic nuclei \citep[e.g.][and reference therein]{amaro-seoane_relativistic_2018}. In this context, rates 
can be obtained via different approaches, perhaps the most popular one being solving the steady-state Fokker-Planck (FP) equation for the distribution function (DF) of the system \citep{2011CQGra..28i4017A,2015ApJ...814...57M,2016ApJ...820..129B}. In the underlying physical model, by setting the inflow of objects in the galactic nucleus the steady-state FP equation is solved to obtain a constant DF. The latter is generally constructed in the energy-angular momentum space, either by integrating over the latter and employing an effective 1D treatment, or by solving the 2D equations. However, as the typical orbital parameters for the production of EMRIs correspond to regions of phase space where the anisotropy of the DF due to the loss-cone is relevant \citep[][]{pan_formation_2021}, the effective 1D treatment of the system \citep[][]{2017ApJ...848...10V} may be inadequate for these phenomena, suggesting that a 2D approach is preferable. Either way, the desired rates are then computed as the number of objects that enter the loss-cone (the region of phase space of captured objects) within the corresponding range of orbital parameters per unit time; this requires an adequate treatment of the boundary condition at the loss-cone interface.

The literature on the subject is vast and a nice summary can be found in \cite{amaro-seoane_relativistic_2018}. As mentioned, FP (and other approximate) models are generally used to describe equilibrium solutions for spherically symmetric systems in the energy-angular momentum space, and sBH capture rates are derived in the steady-state approximation.
This assumption, however, is problematic when transferred to realistic astrophysical systems. Most notably, sBHs capture rates diverge for small MBHs and are inconsistent with the continuous supply of compact objects needed. For example, standard EMRI rates derived with those models are of the order of $10^3\,$Gyr$^{-1}$ for a $10^5\msun$ central MBH, and direct plunge rates are estimated to be at least an order of magnitude higher \citep{2016ApJ...820..129B}. This poses two issues to the steady-state picture: the need of supplying  relatively light MBHs with $10^4$ sBHs per Gyr, and the significant mass growth of the central MBH, which invalidates the fixed central potential assumed in FP calculations. Such high rates, moreover, would exclude existence of intermediate MBHs (below $10^5 {\rm M}_\odot$) for the large mass accretion implied, which makes it difficult to apply them {\it a posteriori} to theoretical MBH population models to compute LISA detection rates. This is currently done by artificially capping the EMRI rate to avoid MBH overgrowth at the faint end of the mass function, as described in \cite{2017PhRvD..95j3012B}. This is obviously unsatisfactory, and a more consistent approach, able to account for the mass growth of the MBH and the finite supply of sBHs, is needed in order to make detailed LISA predictions and to prepare the tools needed to extract those intricate, overlapping signals from the data stream \citep{2010CQGra..27h4009B}. 

The complete FP equation, however, describes the time evolution of a distribution of stellar objects and can be used to compute the rate of capture as a function of time. \cite{pan_formation_2021} solved the time dependent FP equation in two dimensions with a steady potential (\textit{i.e.} the potential is consistent only at start) to simulate larger systems and without assuming a fixed inflow of objects; in this case the injection of compact objects is provided by the migration from far-away orbits to the central region due to the presence of a lighter, dominant, stellar component. The peak in the rates of these systems reproduce the same diverging trend of steady-state counterparts, but the evolution timescale for the systems is so different that the time-average of the capture rates over a fixed time interval is decreasing with the mass of the galaxy for intermediate MBH. This suggests that the time dependent FP approach can be effectively employed to study the detailed evolution of TDEs EMRIs and plunges in more complicated systems, beyond the steady state assumption.


This work is the first in a series of paper aimed at delivering a comprehensive model for the dynamical description of EMRI formation. The final goal is to construct a computationally efficient, time-dependent, two dimensional FP code capable to handle a central potential and a supply of stars and compact objects that are both time-dependent, eventually also including an initial mass function and stellar evolution. Here, we start by developing a two-population (stars and sBHs) FP approach in the two dimensions of energy and angular momentum, to estimate the rates of TDEs, EMRIs and plunges about MBHs in the steady-potential approximation. We discuss the results of several time-dependent FP simulations quantifying the rates of tidal disruptions, direct plunges of compact objects and EMRIs that directly constrain the total growth of the central MBH and its timescale. With these estimates we can identify the limitations of the steady-state model for the EMRI formation rates.
The paper is organized as follows. In Sec.~\ref{sec:gravcap} we introduce the formalism of the loss-cone and the orbit averaged FP equation for nuclear clusters. In Sec.~\ref{sec:algor} we describe our algorithm for solving the FP equation and in Sec.~\ref{sec:sims} we present the results of the various simulations we performed. Finally, we summarise our results and draw our conclusions in Sec.~\ref{sec:concl}.

\section{Gravitational captures}
\label{sec:gravcap}
In this section we briefly review the loss-cone mechanism for gravitational captures, with a focus on their mean--field treatment in the orbit--averaged FP equation \citep{cohn_stellar_1978, merritt_dynamics_2013}.

\subsection{Orbits in the nuclear cluster}
\label{sec:orbits}
We consider a simple model of a galactic nucleus composed by:
\begin{itemize}
    \item a central MBH of mass $M_\bullet$,
    \item a spherical distribution of stars with mass $m_s$,
    \item a spherical, subdominant distribution of stellar mass compact objects with mass $m_{\BH}$.
\end{itemize}
The formalism we use is based on the one used by \cite{pan_formation_2021} and \cite{stone_rates_2016}.
We consider an object -- either a star or an sBH -- orbiting around an MBH located in the centre of a galaxy. Neglecting relativistic corrections, its integrals of motion are the energy per unit mass $E$ and the angular momentum per unit mass $J$, defined by:
\begin{equation}\label{eq:integral_of_motion}
    E = \phi(r) - \frac{v^2}{2} \qquad v^2 = v_r^2 + \frac{J^2}{r^2} 
\end{equation}
where $r$ is the distance from the MBH, $\phi(r)$ is the positive potential of the entire system and $v_r$ is the radial velocity\footnote{Note that with this definition the energy of a bound orbit is positive.}. It is common to rescale the squared angular momentum to its circular-orbit value at energy $E$, introducing the variable
\begin{equation}
\label{eq:J2c}
R = \frac{J^2}{r_c^3 \, \phi'(r_c(E))}
\end{equation}
where  $\phi'(r) = d \phi(r) / dr$, and $r_c(E)$ is the radius of a circular orbit for a test mass with energy $E$, which can be obtained by solving
\begin{equation}
\phi(r) + r\, \frac{\phi'(r)}{2} - E = 0.
\end{equation}
The quantities $E$ and $R$ completely characterise an orbit in a given potential. For example, the radial period of an orbit can be computed as
\begin{equation}\label{eq:orbitalperiod}
    P(E,R) = 2 \int_{r_-}^{r_+} \frac {dr}{v_r}
\end{equation}
where $r_-$ and $r_+$ are the pericentre and the apocentre of the orbit.

Typical stellar densities in nuclear clusters are high enough that mutual interactions between particles can significantly modify the original orbits via \citep{2016ApJ...820..129B}:
\begin{itemize}
    \item[a.] random fluctuations of the orbital parameters, a phenomenon known as \textit{non-resonant relaxation} (NR),
    \item[b.] cumulative non-local effects due to orbit-averaged net torques, which undergo the name of \textit{resonant-relaxation} (RR).
\end{itemize} 
The evolution of the orbital parameters of an object can be described as a Brownian motion in the $(E,J)$ space with fluctuations that depend on the physics of NR and RR, but are in general more pronounced along the $J$ direction, efficiently directing compact objects toward the MBH at the centre of the system \citep{2016ApJ...820..129B}. A compact object like an sBH  can be considered captured once it crosses \citep{merritt_dynamics_2013}
\begin{equation}
    r_\BH = 8 \, \frac{G M_\bullet}{c^2} \,.
    \label{eq:rbh}
\end{equation}
Conversely, an extended object (e.g. a regular star) will be disrupted because of the tidal forces induced by the MBH gravitational field at a distance known as \textit{tidal radius} \citep{stone_rates_2016}, given by
\begin{equation}
\label{eq:rlc_s}
    r_s \simeq r_\star \left(\frac{M_\bullet}{m_s}\right)^{1/3},
\end{equation}
where $r_\star$ is the typical radius of the extended object. At this point, the star is disrupted and a fraction of the debris is captured by the MBH, with the rest escaping on unbound orbits. 

$r_\BH$ and $r_s$ are therefore threshold radii beyond which compact and extended objects are respectively disrupted or captured by the central MBH. In practice, regardless of its initial orbit, any object can be driven below its relevant threshold separation by the cumulative effect of NR and RR. This occurs when the velocity vector of the object is scattered within a small solid angle of the size $r_\BH$ and $r_s$ around the central MBH, which defines a cone-like region in the velocity space, named the 'loss-cone' \citep[because whatever enters this region is 'lost' to the surrounding stellar system; see][for a complete treatment]{amaro-seoane_relativistic_2018}.
In the $(E,R)$ space, the loss-cone is defined by the region below the curve set by the condition on the orbit pericenter
\begin{equation}
    r_-(E, R) \leq r_{\LC},
\end{equation}
where $r_{\LC} = r_{\BH}$ for compact objects and $r_{\LC} = r_s$ for extended objects. This condition gives:
\begin{equation}\label{eq:loss_cone}
    R_{\LC}(E) \leq R \leq 1 \qquad R_{\LC}(E) = 2 \, r^2_{\LC} \, \frac{E - \phi(r_{lc})}{J_c^2(E)}
\end{equation}
for $E < E_{\LC}$, where the latter is the energy of the circular orbit at $r_{\LC}$.
In the next sections we will use the symbols $R_{\LC}^{\BH}$, $E_{\BH}$ and $r_{\BH}$ when considering the gravitational capture of compact objects and $R_{\LC}^s$, $E_s$ and $r_s$ when considering TDEs.

In the case of compact objects we need to distinguish between captures on a direct plunge or an EMRI orbit. The latter occurs when relaxation drives the object on an orbit such that: ($i$) the timescale of the stochastic fluctuations $t_{\rlx}$ becomes comparable to the timescale of the energy loss due to  GWs emission $t_\GW$, but $(ii)$ the pericentre of the object remains larger than $r_\BH$. With good approximation, those two conditions are fullfilled by objects approaching the loss-cone on orbits with energy $E > E_\GW$
\citep{hopman_orbital_2005}, where the latter is a threshold that depends on the potential of the system. Physically, orbits that approach the loss-cone with energy $E_\GW < E < E_\BH$ are dominated by GWs emission: their energy is slowly dissipated until the particle enters the loss-cone at $E\simeq E_\BH$ (remember that we consider a positive definite energy for bound orbits), thus resulting in an EMRI. On the other hand, orbits with $E < E_\GW$ will quickly plunge onto the MBH without significant energy dissipation.


\subsection{Orbit averaged Fokker-Planck equation}
The formation of EMRIs, plunges or TDEs can be considered to be a stochastic process. In the mean field treatment one can write a FP equation for the full 6D distribution function of each component $f^i(\bm x, \bm v)$ ($i$ = \{s, BH\}) in the model. Under the assumption that the potential of the star cluster is dominated by the central MBH and that it evolves on timescales larger than the typical orbital period, it is possible to write the \textit{orbit averaged} FP equation, i.e. an FP equation for the distribution in the space of orbital parameters $(E,R)$:
\begin{equation}\label{eq:FP}
    \frac{\partial}{\partial t}\; \mathcal C(E,R, t) \, f^i(E, R, t) = - \nabla \cdot \bm{\mathcal F_i} (E, R, t)
\end{equation}
where at the left hand side the distribution function $f(E,R,t)$ at time $t$ is multiplied by the weighting function $\mathcal C(E,R,t) = 4\pi^2 \, J_c^2(E) \; P(E,R)$ to give the number density in the $(E,R)$ space. The equation is written in the flux conservation form, where $\bm{\mathcal F}$ is the \textit{current density} in these coordinates. Its general form reads
\begin{equation}\label{eq:density_current}
    \bm{\mathcal F_i} = \begin{pmatrix}\mathcal F_i^E\\ \mathcal F_i^R\end{pmatrix}
    = - \begin{pmatrix}
    \mathcal D_i^{EE} \; \partial_E \, f^i +  \mathcal D_i^{ER} \;  \partial_R \, f^i + \mathcal D_i^E \; f^i \\
    \mathcal D_i^{RE} \;  \partial_E \, f^i +  \mathcal D_i^{RR} \;  \partial_R \, f^i + \mathcal D_i^R \; f^i
    \end{pmatrix}
\end{equation}
where $\mathcal D^E$ and $\mathcal D^R$ are referred to as the \textit{advection coefficients} and the remaining are known as the \textit{diffusion coefficients} of the equation. All these coefficients are functionals of $f(E,R,t)$ and $\phi(r,t)$ and are modeled depending on the underlying physics of the system \citep{merritt_dynamics_2013,2016ApJ...820..129B}.

In \cite{cohn_stellar_1978} and \cite{cohn_numerical_1979} the authors built this formalism for TDEs in a single stellar component system subject to NR only with a boundary condition at the curve defined by equation~\eqref{eq:loss_cone} to effectively treat the loss-cone (we will describe the boundary condition later). The full physical picture with RR and relativistic precession can be included in the modelling of the FP coefficients, as shown by \cite{2016ApJ...820..129B}. In the same work, however, they showed that the steady state rates of the particles across the loss-cone when NR, RR and relativistic precession  are included in the coefficients can be reproduced with acceptable precision by the simpler model including NR only: if particles enter the loss-cone at $E > E_{GW}$ are considered EMRIs, while those entering at $E < E_{GW}$ are considered plunges. In this work we solve the time dependent equation \eqref{eq:FP} with NR only, assuming that the rates obtained are indicative of the full physical framework as in the steady state case.

The value of $E_{\rm GW}$ depends on the masses of the stellar compact object ($m_\BH$) and the MBH ($M_\bullet$), as well as on the properties of the potential \citep{hopman_orbital_2005}. An estimate of the semi-major axis of the orbit at $E_\GW$ is
\begin{equation}
    r_{\GW} \sim \left[\frac{0.035}{m_{\BH}} \frac{M_\bullet}{N_h} \frac{1}{\log \frac{M_\bullet}{m_{\BH}} + \frac 14 \log \frac{2 \sigma^2}{c^2}} \right]^{1/\gamma}\; r_h
\end{equation}
where
\begin{equation}
r_h=\frac{GM_\bullet}{\sigma^2} 
\end{equation}
is the MBH influence radius, $N_h$ is the number of stars inside $r_h$, $\sigma$ is the velocity dispersion of the stellar bulge and  $-\gamma$ is the inner slope of its density profile. We explore different values of $M_\bullet$ and $\sigma$ with $\gamma = 1.5$, as detailed in section~\ref{sec:simpar}, and the ratio $r_{\GW}/r_h$ ranges between $0.005$ to $0.014$; for this reason we opted for the value commonly used in literature $r_{\GW} = 0.01 r_h$. This approximation remains valid also for slightly $\gamma = 1.2,\ 1.8$.

\section{Algorithm}
\label{sec:algor}

In this section we will present the initial conditions for the simulated systems and the algorithm used to perform the simulation. The general framework is based on the work of \cite{pan_formation_2021}.

\subsection{Initial conditions}
 We fix the total mass of extended objects (stars hereinafter) in the distribution to $M_s = 20 \, M_\bullet$, composed by stars of mass $m_s = 1 \, M_\odot$ each, and the total mass of compact objects (sBHs hereinafter) to $M_{\BH} = 0.2 \, M_\bullet$, composed by sBHs of $m_{\BH} = 10 \, M_\odot$ each.

Both distributions initially follow the same Dehnen profile with $\gamma = 1.5$ \citep{binney_galactic_2009}
\begin{equation}\label{eq:numberdensity}
    n_i(r) = \frac{3}{8\pi} \; N_i \; \frac{r_a}{r^{3/2} \; (r + r_a)^{5/2}} \, \theta(r - r_i).
\end{equation}
Here $r_a$ sets the radius at which the distribution changes its behaviour from $r^{-1.5}$ to $r^4$ and the Heaviside function $\theta$ is introduced to truncate the distribution at $r=r_i$, where $r_i=r_\BH$ for sBHs and $r_i=r_s$ for stars. In order to adequately scale $r_a$ with the properties of the system, we set it to a multiple of the influence radius of the MBH $r_h$ \citep{merritt_dynamics_2013}
\begin{equation}
    r_a = 4 \, r_h.
\end{equation}
Then, we use the $M_\bullet - \sigma$ relation \citep{gultekin_them-upsigma_2009} to finally set the scale radius of the distribution
\begin{equation}\label{eq:Msigma}
    \sigma = 70\ \textrm{Km/s} \cdot \left ( \frac{M_\bullet }{ 1.53 \  10^6 M_\odot} \right)^{1/4.24} \, .
\end{equation}
The ratio $r_a / r_h$ is somewhat arbitrary, since the velocity dispersion in the central region in our model is not constant and diverges as $\sim r^{-1}$ at the centre. For a comparison, the stellar radial velocity dispersion of the initial conditions can be computed (see Appendix \ref{ap:stellarsigma}). Neglecting the sBHs contribution we obtain
\begin{equation}
    \sigma_r^2(r_h) \simeq 4.6 \, \frac{G \, M_\bullet}{r_a} \, , \qquad  \sigma_r^2(r_a) \simeq 2.4 \, \frac{G \, M_\bullet}{r_a}
\end{equation}
and even more sophisticated estimates -- \textit{e.g.} averages in the central regions -- give similar numerical factors. For an MBH of $M_\bullet = 4 \times 10^6 \rm M_\odot$ our model has a density of $6\times 10^4 \rm M_\odot/pc^3$ at the influence radius $r_h = 2.23$ pc, which is consistent with observational estimates for the Milky Way \citep[e.g.][]{2007A&A...469..125S}. The total potential of the MBH plus the extended star and sBH nuclear cluster is
\begin{equation}
    \phi(r) = \frac{GM_\bullet}{r} + 2 \, G (M_s + M_{\BH}) \, \left(1 - \sqrt{\frac{r}{r + r_a}}\, \right)
\end{equation}
and is not evolved during the simulation. The initial distribution function $f^i(E,R) $ of stars and sBHs is set by computing the isotropic distribution function of equation~\eqref{eq:numberdensity} using the Abel transform \citep{binney_galactic_2009}

\begin{equation}\label{eq:Initial_f_0}
f^i(E,R,t=0) = \frac{\sqrt{2}}{4 \, \pi^2} \frac{d}{dE} \int_0^E d\phi \; \frac{d n_i(r(\phi))}{d\phi} \; \frac{1}{\sqrt{E - \phi}}\,\,,
\end{equation}
and restricting only to the phase space region of bound orbits outside the loss-cone.

In order to numerically integrate equation \eqref{eq:FP} we represent the distribution function on a uniform grid in the variables $(s,R)$, where $s$ reads
\begin{equation}
    s = \log \left( 1 + \frac{E}{E_0}\right)\,.
\end{equation}
Here $E_0$ is a reference energy scale set to $\sigma^2/5$ in the simulations. Due to the loss-cone, the support of $f^i$ is naturally compact and is a subregion of the square domain $s \in [0, s_{lc}]$ and $R \in [0,1]$.
To numerically compute $f^i(s,R,0)$ we rearrange equation \eqref{eq:Initial_f_0} in the form
\begin{equation}
\label{eq:Initial_f_2}
f^i(s, R, 0) = \frac{\sqrt{2 \, E_0}}{4\pi^2}  \; \int_0^s dw\,\frac{e^w }{\sqrt{e^s - e^w}} \;\frac{d^2\,n^i}{d\phi^2}
\end{equation}
and we use Gauss-Legendre quadrature to compute the function at the grid values of $s$. We represent the distribution function of each component on a uniform grid $N_s \times N_R$.

\subsection{Coefficients computation and boundary conditions}
The complete expressions for the coefficients and the auxiliary functions needed for their computation are reported in Appendix \ref{ap:FP_coeffs}. Here we  schematically report the set of equations and steps needed for their computation in our numerical approach:
\begin{enumerate}
    \item marginalise $f^i$ over $R$
    \begin{equation}
         \bar f^i(s) = \int_0^1 dR \, f(s, R) 
    \end{equation}
    and build a linear interpolant of the function;
    \item compute a set of auxiliary functions that depend on the variable $w = \log(1 + \phi(r)/E_0)$
    \begin{equation}
        F_i^k(s, w) = \int_s^w ds' \, h^k(s, w, s') \, \bar f^i (s') 
    \end{equation}
    where $h^k(s, w, s') $ is a smooth and compact weighting function. We compute these integrals at a uniform grid $s \times w$; the values of $s$ are the same as those of the grid of $f^i$ and the values of $w$ have the same spacing of $s$, covering the domain $[0, w(r_\LC)]$. The integration is performed with the Gauss-Legendre quadrature technique;

    \item compute the FP coefficients
    \begin{equation}
        \mathcal D_i^j(s, R) = \int_{w_+}^{w_-} dw \, \sum^i_{j,k} \; \mu_{i,k}^{j}(s, R, w) \, F_i^k(s, w)
    \end{equation}
    on the original grid of $f^i$. The weights $\mu^j_{i,k}$ have a divergent denominator proportional to $v_r$ that behaves like $\sqrt{|w-w_{-/+}|}$ at the endpoints. The divergences have been treated using the Gauss-Chebyshev quadrature technique after explicit extraction. 
\end{enumerate}
For both point \textit{(ii)} and point \textit{(iii)} we used 300 nodes for Gauss quadrature. Since the potential is fixed, for a given point of the grid of $f$ the value of the weights $\mu^i_{jk}$ at the Chebyshev points of $[w_-, w_+]$ does not change during the evolution; we compute them in advance to make the computation of the coefficients faster.

The boundary conditions for each component are:
\begin{itemize}
    \item At $s = 0$ all the coefficients vanish and the function does not evolve with time. In our case this fixes the value $f^i(0, R) = 0$ at all times.
    \item The line $R=1$ is the locus of circular orbits. Since at fixed energy a particle cannot exceed $J^2_c(E)$, the flux along $R$ vanishes by construction for circular orbits and all the $\mathcal D^i_{Rx}$ coefficients vanish.
    \item At the loss-cone boundary the dominant coefficient is $\mathcal D^i_{RR}$ and one can compute the boundary behaviour of $f^i(s,R)$ (and, in particular, of its derivative) at fixed $s$ \citep{cohn_stellar_1978, merritt_dynamics_2013}
    \begin{equation}
    \label{eq:loss_cone_behaviour}
     f^i \approx f(R^i_\LC) \left[ 1 + \frac{\ln R/R^i_\LC}{\ln R^i_\LC / R^i_0}\right]
    \qquad \textrm{for } R \to R^i_{\LC}\end{equation}
    where $R_0$ is given by the following approximate relation
    \begin{equation}
     R^i_0 \simeq R^i_{\LC}\; \exp \left( -\sqrt[4]{q_i^4 + q_i^2} \right)
    \end{equation}
    and
    \begin{equation}
     q_i = \frac{1}{4 \, \pi^2 J^2_c\, R^i_{\LC}(s_i)}\, \lim_{R \to R^{i \, +}_{\LC}} \frac{\mathcal D^{RR}_i}{R} \, .
    \end{equation}
    The limit at a given value of $s$ is numerically performed by evaluating the quantity $q$ at the first grid point above the loss-cone curve.
\end{itemize}

\section{Simulations}
\label{sec:sims}

In this section we present the results of the simulations that we have run leveraging on the Julia \citep{bezanson_julia_2017} implementation of the algorithm described in the previous section\footnote{We plan to publicly release the code together with an extensive description of the numerical implementation in a forthcoming paper. The interested reader can find a first version of the code at the following repository: \url{https://gitlab.com/j2970/juliafokkerplanck}}.

\subsection{Performed simulations}\label{sec:simpar}
\begin{figure}
    \centering
    \includegraphics[width=\linewidth]{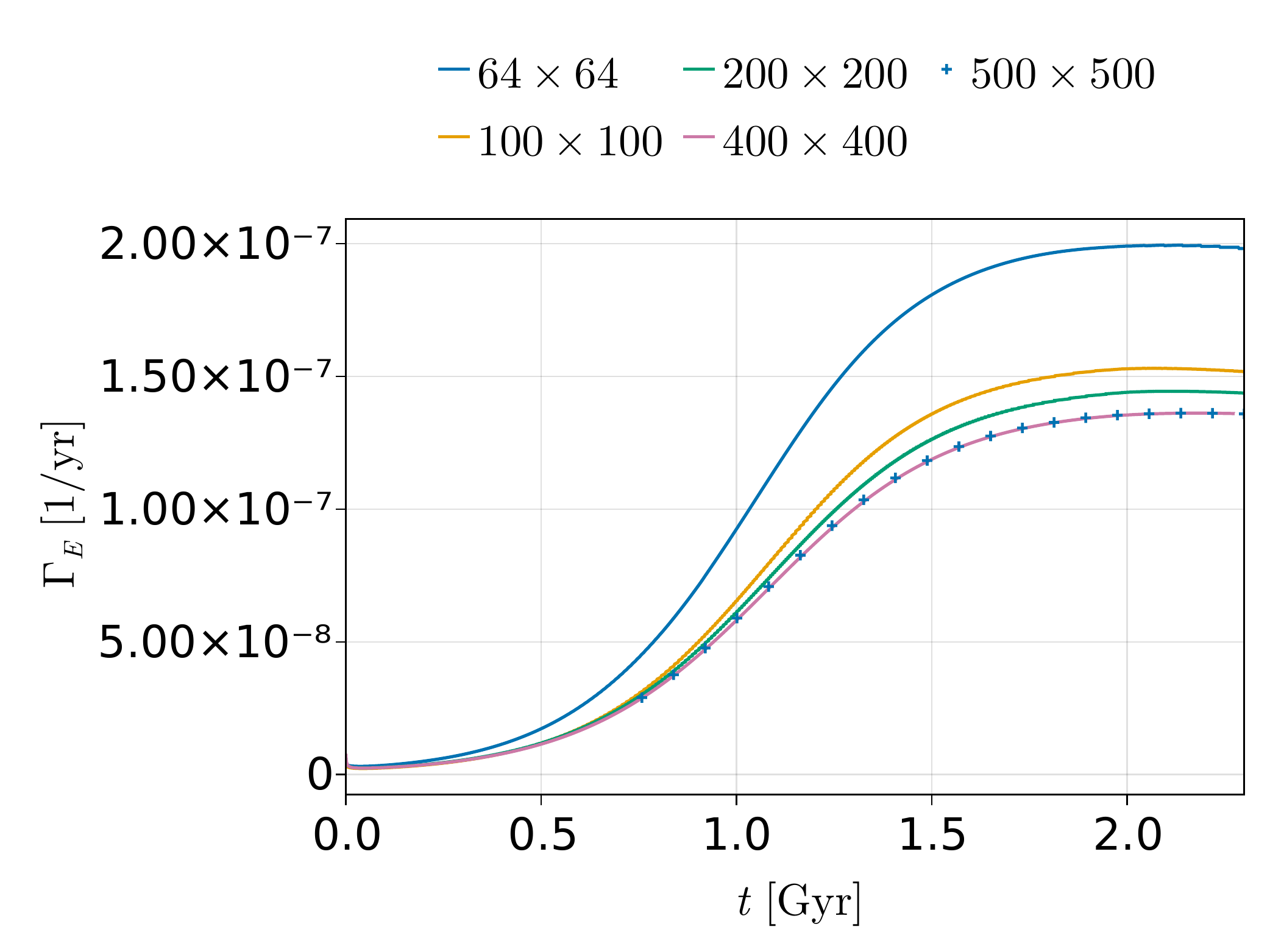}\\
    \includegraphics[width=\linewidth]{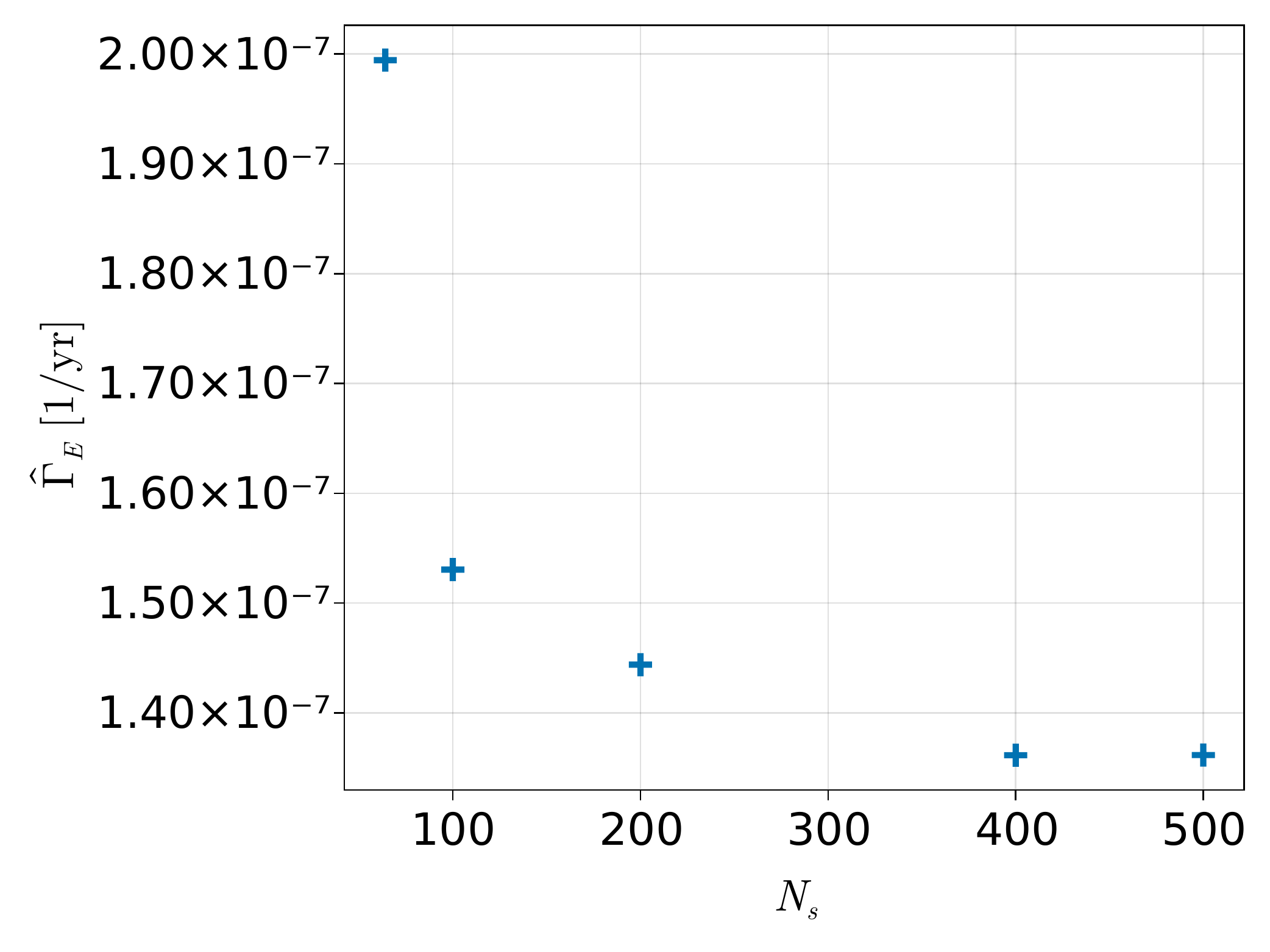}
    \caption{Top panel: EMRI formation rate for the same simulation at different resolutions with $N_R = N_s$. The system simulated has $M_\bullet = 4 \cdot 10^6 \, M_\odot$ with $\sigma$ taken from the $M_\bullet - \sigma$ relation. The highest resolution is represented as a point series for better readability. Bottom panel: the maximum of each curve as a function of $N_s$. In the rest of this work we set $N_R = N_s = 200$.}
    \label{fig:resolution}
\end{figure}

We considered seven values of $M_\bullet$ evenly spaced in log scale in the range  $10^4 M_\odot$ -- $10^7 M_\odot$.
For each $M_\bullet$ we computed the density $\rho_0$ corresponding to the value of $\sigma_0$ obtained from \eqref{eq:Msigma} and consider the values $\rho = \{0.01,\ 0.1,\ 1,\ 10\} \ \rho_0$ (which correspond to $\sigma = \{0.464,\ 0.681,\ 1,\ 1.468\} \ \sigma_0$), for a total of 28 simulations.


We performed convergence tests simulating a Milky Way like system with $M_\bullet = 4\times 10^6\, M_\odot$ using grids of different resolutions (see Fig. \ref{fig:resolution}) and we opted for $N_s\times N_R = 200\times200$ to balance accuracy and computational time. 

The time step is set by limiting the maximum relative variation of the DF on the grid
\begin{equation}
    \max_{grid} \frac{\Delta f}{f} = 0.03.
\end{equation}

We run each simulation for a total time
\begin{equation}
    t_f = \min \left(10 \, t_E,\ 10\ \mathrm{Gyr}\right)
\end{equation}
where $t_E$ is the time when the EMRI rate peaks (see section~\ref{sec:rates} for details) and 10 Gyr has been chosen as representative of the Hubble time. 

\begin{landscape}
\begin{table}
\centering
\begin{tabular}{rrr|rrr|rrr|rrr|rrr}
 $M_{\BH}$ $[M_\odot]$ & $r_h$ [pc] & $\rho_s(r_h)$ [$M_\odot$/pc$^3$] & $\hat \Gamma_{E}$ [1/yr] & $t_{E}$ [Gyr]& $N_{E}$ & $\hat \Gamma_{P}$ [1/yr] & $t_{P}$ [Gyr] & $N_{P}$ & $\hat \Gamma_{T}$ [1/yr] & $t_{T}$ [Gyr] & $N_{T}$ & $t_{\bullet}$ [Gyr] & $t_{f}$ [Gyr]& $\tau$\\ \hline
$ 10^{4}$ & $0.44$ & $2.0 \times 10^{4}$ & $5.8 \times 10^{-8}$ & $0.089$ & $2.6$ & $5.2 \times 10^{-7}$ & $0.017$ & $37$ & $1.4 \times 10^{-4}$ & $0.0036$ & $6.7 \times 10^{3}$ & $0.23$ &$0.090$& $10$\\
$ 10^{4}$ & $0.2$ & $2.0 \times 10^{5}$ & $1.9 \times 10^{-7}$ & $0.0029$ & $2.7$ & $1.6 \times 10^{-6}$ & $0.0054$ & $38$ & $4.4 \times 10^{-4}$ & $0.0011$ & $7.3 \times 10^{3}$ & $0.071$ &$0.029$& $10$\\
$ 10^{4}$ & $0.094$ & $2.0 \times 10^{6}$ & $6.0 \times 10^{-7}$ & $0.00092$ & $2.8$ & $5.1 \times 10^{-6}$ & $0.0017$ & $38$ & $0.0015$ & $5.3 \times 10^{-4}$ & $7.5 \times 10^{3}$ & $0.022$ &$0.0092$& $10$\\
$ 10^{4}$ & $0.44$ & $2.0 \times 10^{7}$ & $2.0 \times 10^{-6}$ & $0.00030$ & $3.0$ & $1.6 \times 10^{-5}$ & $5.5 \times 10^{-4}$ & $38$ & $0.0047$ & $2.0 \times 10^{-4}$ & $7.9 \times 10^{3}$ & $0.0068$ &$0.0030$& $10$\\
$3 \times 10^{4}$ & $0.78$ & $1.1 \times 10^{4}$ & $4.5 \times 10^{-8}$ & $0.037$ & $8.5$ & $3.8 \times 10^{-7}$ & $0.070$ & $1.1 \times 10^{2}$ & $1.0 \times 10^{-4}$ & $0.015$ & $2.1 \times 10^{4}$ & $0.93$ &$0.37$& $10$\\
$3 \times 10^{4}$ & $0.36$ & $1.1 \times 10^{5}$ & $1.4 \times 10^{-7}$ & $0.012$ & $8.3$ & $1.2 \times 10^{-6}$ & $0.022$ & $1.1 \times 10^{2}$ & $3.3 \times 10^{-4}$ & $0.0048$ & $2.2 \times 10^{4}$ & $0.29$ &$0.12$& $10$\\
$3 \times 10^{4}$ & $0.17$ & $1.1 \times 10^{6}$ & $4.7 \times 10^{-7}$ & $0.0038$ & $9.0$ & $3.7 \times 10^{-6}$ & $0.0070$ & $1.1 \times 10^{2}$ & $0.0011$ & $0.0015$ & $2.3 \times 10^{4}$ & $0.90$ &$0.038$& $10$\\
$3 \times 10^{4}$ & $0.78$ & $1.1 \times 10^{7}$ & $1.5 \times 10^{-6}$ & $0.0012$ & $9.4$ & $1.2 \times 10^{-5}$ & $0.0022$ & $1.1 \times 10^{2}$ & $0.0035$ & $8.1 \times 10^{-4}$ & $2.4 \times 10^{4}$ & $0.028$ &$0.012$& $10$\\
$ 10^{5}$ & $1.5$ & $5.3 \times 10^{3}$ & $3.2 \times 10^{-8}$ & $0.18$ & $29$ & $2.7 \times 10^{-7}$ & $0.33$ & $3.8 \times 10^{2}$ & $7.2 \times 10^{-5}$ & $0.072$ & $7.2 \times 10^{4}$ & $4.4$ &$1.77$& $10$\\
$ 10^{5}$ & $0.69$ & $5.3 \times 10^{4}$ & $1.0 \times 10^{-7}$ & $0.056$ & $29$ & $8.4 \times 10^{-7}$ & $0.10$ & $3.8 \times 10^{2}$ & $2.3 \times 10^{-4}$ & $0.023$ & $7.4 \times 10^{4}$ & $1.4$ &$0.56$& $10$\\
$ 10^{5}$ & $0.32$ & $5.3 \times 10^{5}$ & $3.3 \times 10^{-7}$ & $0.018$ & $30$ & $2.7 \times 10^{-6}$ & $0.033$ & $3.8 \times 10^{2}$ & $7.8 \times 10^{-4}$ & $0.0072$ & $7.7 \times 10^{4}$ & $0.42$ &$0.18$& $10$\\
$ 10^{5}$ & $0.15$ & $5.3 \times 10^{6}$ & $1.1 \times 10^{-6}$ & $0.0057$ & $32$ & $8.4 \times 10^{-6}$ & $0.011$ & $3.8 \times 10^{2}$ & $0.0026$ & $0.0023$ & $9.0 \times 10^{4}$ & $0.13$ &$0.57$& $10$\\
$3 \times 10^{5}$ & $2.6$ & $2.8 \times 10^{3}$ & $2.4 \times 10^{-8}$ & $0.73$ & $88$ & $2.0 \times 10^{-7}$ & $1.4$ & $1.1 \times 10^{3}$ & $5.2 \times 10^{-5}$ & $0.3$ & $2.1 \times 10^{5}$ & $18$ &$7.3$& $10$\\
$3 \times 10^{5}$ & $1.2$ & $2.8 \times 10^{4}$ & $7.9 \times 10^{-8}$ & $0.23$ & $90$ & $6.2 \times 10^{-7}$ & $0.43$ & $1.1 \times 10^{3}$ & $1.7 \times 10^{-4}$ & $0.095$ & $2.2 \times 10^{5}$ & $5.7$ &$2.3$& $10$\\
$3 \times 10^{5}$ & $0.57$ & $2.8 \times 10^{5}$ & $2.6 \times 10^{-7}$ & $0.075$ & $97$ & $1.9 \times 10^{-6}$ & $0.14$ & $1.2 \times 10^{3}$ & $5.6 \times 10^{-4}$ & $0.030$ & $2.3 \times 10^{5}$ & $1.8$ &$0.75$& $10$\\
$3 \times 10^{5}$ & $0.26$ & $2.8 \times 10^{6}$ & $8.7 \times 10^{-7}$ & $0.023$ & $1.0 \times 10^{2}$ & $6.1 \times 10^{-6}$ & $0.044$ & $1.2 \times 10^{3}$ & $0.0019$ & $0.0095$ & $2.4 \times 10^{5}$ & $0.54$ &$0.24$& $10$\\
$ 10^{6}$ & $5$ & $1.4 \times 10^{3}$ & $1.8 \times 10^{-8}$ & $3.5$ & $1.8 \times 10^{2}$ & $1.4 \times 10^{-7}$ & $6.5$ & $1.1 \times 10^{3}$ & $3.6 \times 10^{-5}$ & $1.43$ & $3.0 \times 10^{5}$ & $61$ &$10$& $2.9$\\
$ 10^{6}$ & $2.3$ & $1.4 \times 10^{4}$ & $6.0 \times 10^{-8}$ & $1.1$ & $3.1\times 10^{2}$ & $4.4 \times 10^{-7}$ & $2.1$ & $3.5 \times 10^{3}$ & $1.2 \times 10^{-4}$ & $0.46$ & $6.9 \times 10^{5}$ & $26$ &$10$& $9.1$\\
$ 10^{6}$ & $1.1$ & $1.4 \times 10^{5}$ & $2.0 \times 10^{-7}$ & $0.35$ & $3.5 \times 10^{2}$ & $1.4 \times 10^{-6}$ & $0.65$ & $3.8 \times 10^{3}$ & $3.9 \times 10^{-4}$ & $0.14$ & $7.5 \times 10^{5}$ & $8.3$ &$3.46$& $10$\\
$ 10^{6}$ & $0.5$ & $1.4 \times 10^{6}$ & $6.8 \times 10^{-7}$ & $0.11$ & $3.7 \times 10^{2}$ & $4.3 \times 10^{-6}$ & $0.21$ & $3.8 \times 10^{3}$ & $0.0013$ & $0.45$ & $7.8 \times 10^{5}$ & $2.6$ &$1.1$& $10$\\
$3 \times 10^{6}$ & $8.9$ & $7.3 \times 10^{2}$ & $1.2 \times 10^{-8}$ & $10$ & $41$ & $7.5 \times 10^{-8}$ & $10$ & $4.2 \times 10^{2}$ & $2.6 \times 10^{-5}$ & $5.8$ & $2.6 \times 10^{5}$ & $2.2 \times 10^{2}$ $10$&$10$& $1$\\
$3 \times 10^{6}$ & $4.1$ & $7.2 \times 10^{3}$ & $4.6 \times 10^{-8}$ & $4.6$ & $3.3 \times 10^{2}$ & $3.2 \times 10^{-7}$ & $8.5$ & $2.5 \times 10^{3}$ & $8.5 \times 10^{-5}$ & $1.84$ & $7.5 \times 10^{5}$ & $74$ &$10$& $2.2$\\
$3 \times 10^{6}$ & $1.9$ & $7.3 \times 10^{4}$ & $1.6 \times 10^{-7}$ & $1.5$ & $9.3 \times 10^{3}$ & $1.0 \times 10^{-6}$ & $2.7$ & $8.5 \times 10^{3}$ & $2.8 \times 10^{-4}$ & $0.60$ & $1.8 \times 10^{6}$ & $30$ &$10$& $6.9$\\
$3 \times 10^{6}$ & $0.89$ & $7.3 \times 10^{5}$ & $5.6 \times 10^{-7}$ & $0.46$ & $1.3 \times 10^{3}$ & $3.2 \times 10^{-6}$ & $0.85$ & $1.2 \times 10^{4}$ & $9.2 \times 10^{-4}$ & $0.19$ & $2.3 \times 10^{6}$ & $11$ &$4.6$& $10$\\
$ 10^{7}$ & $17$ & $3.6 \times 10^{2}$ & $4.0 \times 10^{-10}$ & $10$ & $2.4$ & $1.9 \times 10^{-8}$ & $10$ & $1.8 \times 10^{2}$ & $1.8 \times 10^{-5}$ & $9.9$ & $2.0 \times 10^{5}$ & $1.0 \times 10^{3}$ $10$&$0.090$& $1$\\
$ 10^{7}$ & $7.8$ & $3.6 \times 10^{3}$ & $1.4 \times 10^{-8}$ & $10$ & $41$ & $1.1 \times 10^{-7}$ & $10$ & $7.6 10^{2}$ & $5.9 \times 10^{-5}$ & $9.0$ & $6.0 \times 10^{5}$ & $3.2 \times 10^{2}$ $10$&$0100$& $1$\\
$ 10^{7}$ & $3.6$ & $3.6 \times 10^{4}$ & $1.2 \times 10^{-7}$ & $6.9$ & $8.1 \times 10^{2}$ & $7.0 \times 10^{-7}$ & $10$ & $4.9 \times 10^{3}$ & $1.9 \times 10^{-4}$ & $2.7$ & $1.8 \times 10^{6}$ & $1.0 \times 10^2$ &$10$& $1.4$\\
$ 10^{7}$ & $1.7$ & $3.6 \times 10^{5}$ & $4.4 \times 10^{-7}$ & $2.1$ & $2.8 \times 10^{3}$ & $2.2 \times 10^{-6}$ & $4.0$ & $2.0 \times 10^{4}$ & $6.4 \times 10^{-4}$ & $0.87$ & $4.6 \times 10^{6}$ & $40$ &$10$& $4.7$\\
\end{tabular}
\caption{\label{tbl:runs}Table summarising the results of all the runs. For each simulation we report the physical parameters (MBH mass $M_\bullet$, influence radius $r_h$ and the density of stars $\rho(r_h)$); for the various rates $\Gamma_x$ we report the critical point value, the time $t_x$ when the value is reached and the total number of events $N_x$ in the simulation. Finally, we report the typical time of growth $t_\bullet$, the total time of the simulation $t_f$ and the ratio $\tau = t_f/t_E$. The grid resolution is $N_R\times N_s$ = $200\times200$.}
\end{table}

\begin{table}
    \centering
\begin{tabular}{rrr|rrr|rrr|rrr|rrr}
$\gamma$ & $r_h$ [pc] & $\rho_s(r_h)$ [$M_\odot$/pc$^3$] & $\hat \Gamma_{E}$ [1/yr] & $t_{E}$ [Gyr]& $N_{E}$ & $\hat \Gamma_{P}$ [1/yr] & $t_{P}$ [Gyr] & $N_{P}$ & $\hat \Gamma_{T}$ [1/yr] & $t_{T}$ [Gyr] & $N_{T}$ & $t_{\bullet}$ [Gyr] & $t_{f}$ [Gyr]& $\tau$\\ \hline
$1.2$& $2.2$ & $4.6 \times 10^{4}$ & $5.2 \times 10^{-8}$ & $5.6$ & $1.3 \times 10^{3}$ & $6.0 \times 10^{-7}$ & $8.7$ & $2.4 \times 10^{4}$ & $1.2 \times 10^{-4}$ & $3.2$ & $3.9 \times 10^{6}$ & $1.0 \times 10^2$ &$57$& $10$\\
$1.5$ & $2.2$ & $6.1 \times 10^{4}$ & $1.5 \times 10^{-7}$ & $2.1$ & $1.5 \times 10^{3}$ & $9.2 \times 10^{-7}$ & $3.9$ & $1.5 \times 10^{4}$ & $2.6 \times 10^{-4}$ & $0.93$ & $3.0 \times 10^{6}$ & $51$ &$21$& $10$\\
$1.8$ & $2.2$ & $7.9 \times 10^{4}$ & $7.0 \times 10^{-7}$ & $0.44$ & $1.8 \times 10^{3}$ & $1.7 \times 10^{-6}$ & $1.0$ & $6.6 \times 10^{3}$ & $9.5 \times 10^{-4}$ & $0.05$ & $1.9 \times 10^{6}$ & $17$ &$4.4$& $10$\\
\end{tabular}
    \caption{Data relative to the simulations with $M_\bullet = 4\times10^6\, M_\odot$ and different central slope of the Dehnen profile $\gamma$. The grid resolution is $N_R\times N_s$ = $200\times200$.}
    \label{tab:various_gamam}
\end{table}

\end{landscape}

\begin{figure*}
    \centering
    \includegraphics[width=\linewidth]{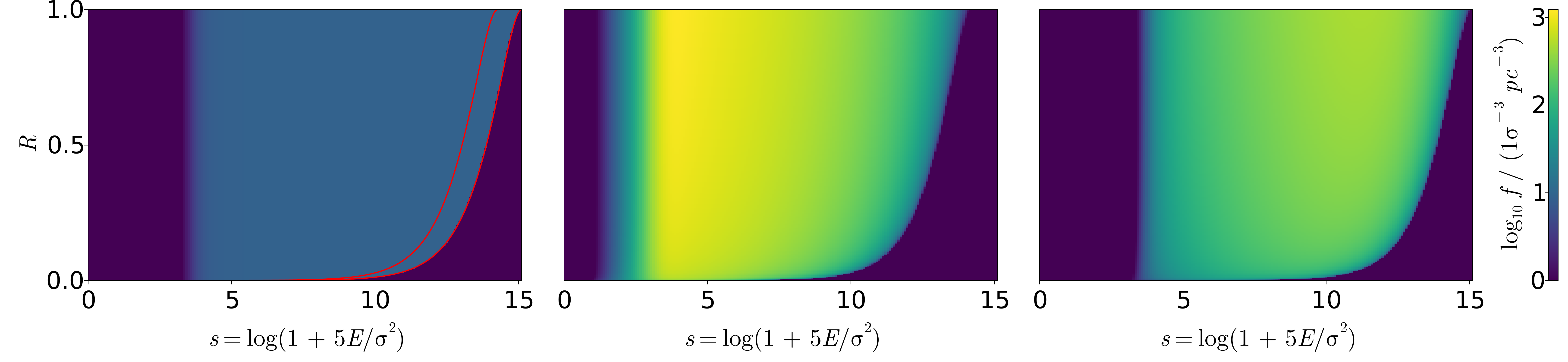}
    \caption{Distribution functions for a system with $M_\bullet = 4 \times 10^6 \, M_\odot$ and stellar objects distributed as a Dehnen profile with $\gamma = 1.5$ and scale radius $r_a = 4 \, r_h$. The left panel shows the initial distribution of sBHs; the initial distribution of stars is larger by a factor of $1000$ and vanishes below the stellar loss-cone (upper red line). The middle and the right panels show the distribution of stars and sBHs respectively at $t = 10$ Gyr. The maps show the effects of mass segregation (see the main text for details).}
    \label{fig:DF}
\end{figure*}

\begin{figure}
    \centering
    \includegraphics[width=\linewidth]{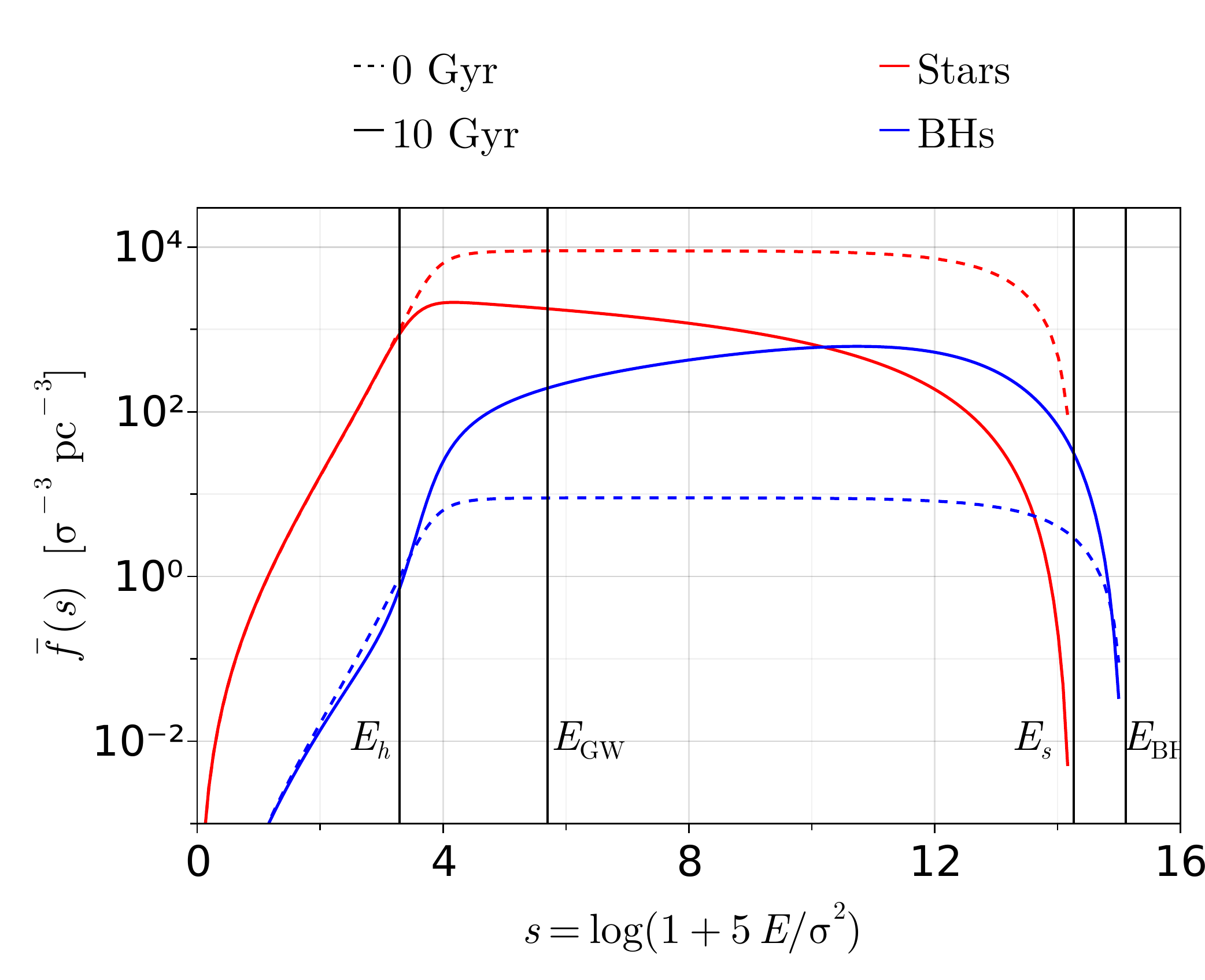}
    \caption{Distribution function $\bar{f}(s)$ marginalised over $R$ for the two components at the beginning (0 Gyr) and at the end of the simulation (10 Gyr) for the DFs in Fig.~\ref{fig:DF}. The vertical solid black line at $E_x$ corresponds to the circular orbit at $r_x$; we consider the influence radius $r_h$, the EMRIs-plunges delimiter $r_\GW$, and the capture raidus for the components $r_s$ and $r_\BH$}
    \label{fig:fbar}
\end{figure}

\begin{figure}
    \centering
    \includegraphics[width=\linewidth]{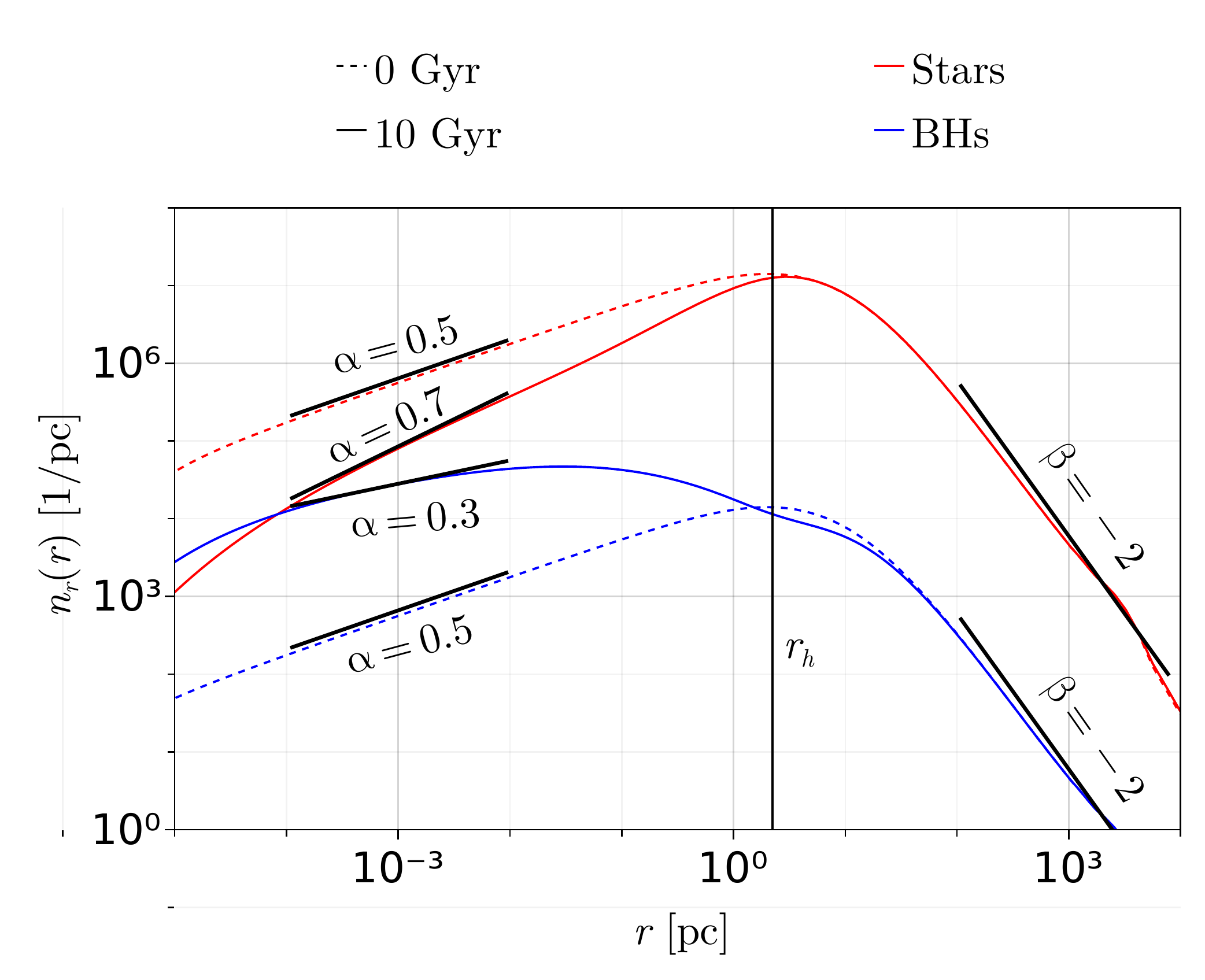}
    \caption{Initial and final radial distribution of stars and sBHs at $t=0$ Gyr and $t=10$ Gyr for the DFs in Fig.~\ref{fig:DF}. The parameters of the simulation are $M_\bullet = 4\times10^6 \, M_\odot$, $N_R = N_s = 200$. We report a power-law fit in the inner (index $\alpha$) and outer regions (index $\beta$).}
    \label{fig:mdistrib}
\end{figure}

\begin{figure}
    \centering
    \includegraphics[width=\linewidth]{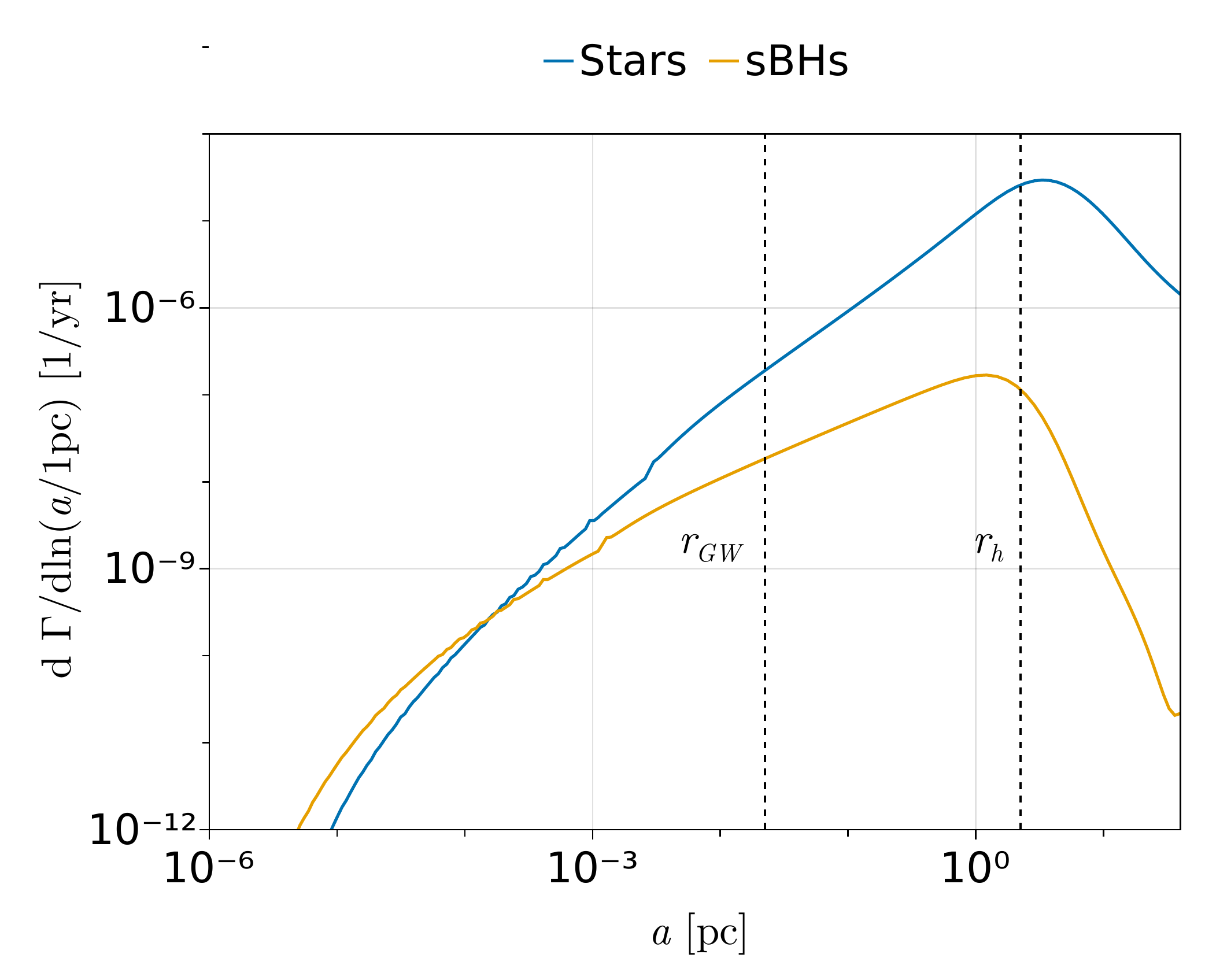}
    \caption{Flux of each of the two considered species entering the loss-cone as a function of the semimajor axis at $10$ Gyr for a system with $M_\bullet=4\times10^6 M_\odot$ on the $M_\bullet - \sigma$ (the correspondent DFs are shown in the central and right panels of Fig.~\ref{fig:DF}).}
    \label{fig:enteringflux}
\end{figure}

\subsection{Final state}
In Fig. \ref{fig:DF} we show the DF at the beginning (t=0 Gyr) and at the end (t=10 Gyr) of the simulation for our model with $M_\bullet=4\times10^6 \, M_\odot$. The color maps highlight the effect of mass segregation, which pushes the BHs towards high values of $s$ while relegating stars at small $s$, far from the central object.
This is more evident when marginalizing the DF over the angular momentum variable $R$, as shown in Fig. \ref{fig:fbar}. At the end of the simulation the sBHs distribution is more concentrated at circular orbits and higher energies, located at smaller distances from the centre. In Fig. \ref{fig:mdistrib} we show the radial distribution of the two components in our system at $t=0$ Gyr and $t=10$ Gyr. The lighter stellar component increases the central slope from $\alpha = 0.5$ to $\alpha = 0.7$, while the heavier sBHs component shows a milder behaviour ($\alpha \simeq 0.3$ at $10^3 \, r_{b}~\simeq 1$~mpc). These slopes correspond to a 3D particle density that behaves like $n\propto r^{-\gamma}$ with $\gamma=1.3$ for the light component and $\gamma = 1.7$ for the heavy one.
This mass segregation effect is due to the fact that 2-body interactions explicitly depend on the mass of the components. In the FP equation, mass segregation is encoded in the coefficients: while diffusion coefficients are identical for the two components, advection coefficients are proportional to the mass $m_i$\footnote{Note that, in realistic stellar systems, the effect of supernova explosions may render mass segregation even more pronounced 
\citep{2017MNRAS.469.1510B};  at the same time, as mentioned in the introduction, those events can represent a further mechanism to generate EMRIs and plunges \citep{2019MNRAS.485.2125B, 2022arXiv220403661H}}. The inner slopes of stars and sBHs we find at the end of our integrations are slightly shallower than the theoretical expectations based on the 1D FP equation by \citet{1977ApJ...216..883B}. The presence of the loss-cone requires a non-zero (positive) flux towards the central regions, compatible with lower power-law indices. Moreover, the strong anisotropy of the DF in the $E \to E_\LC$ region may contribute to the flux via the cross-term, which is neglected in the 1D equation.

In Fig. \ref{fig:enteringflux} we plot the rate across the loss-cone at $t=10$ Gyr with $M_\bullet=4 \times 10^6 M_\odot$. Most of the particles are captured by the MBH around the influence radius. This is a direct consequence of the higher efficiency of diffusion in $R$ rather then $E$. As a consequence, objects are preferentially scattered into the loss-cone orbit at very large separations, of the order of $r_h$, and captured by the central BH onto very eccentric orbits.    
The effect of mass segregation is also evident here: the stellar rate within loss-cone is steeper than the sBHs rate since the latter are driven towards the MBH.

\subsection{Rates}
\label{sec:rates}
\begin{figure}
    \centering
    \includegraphics[width=\linewidth]{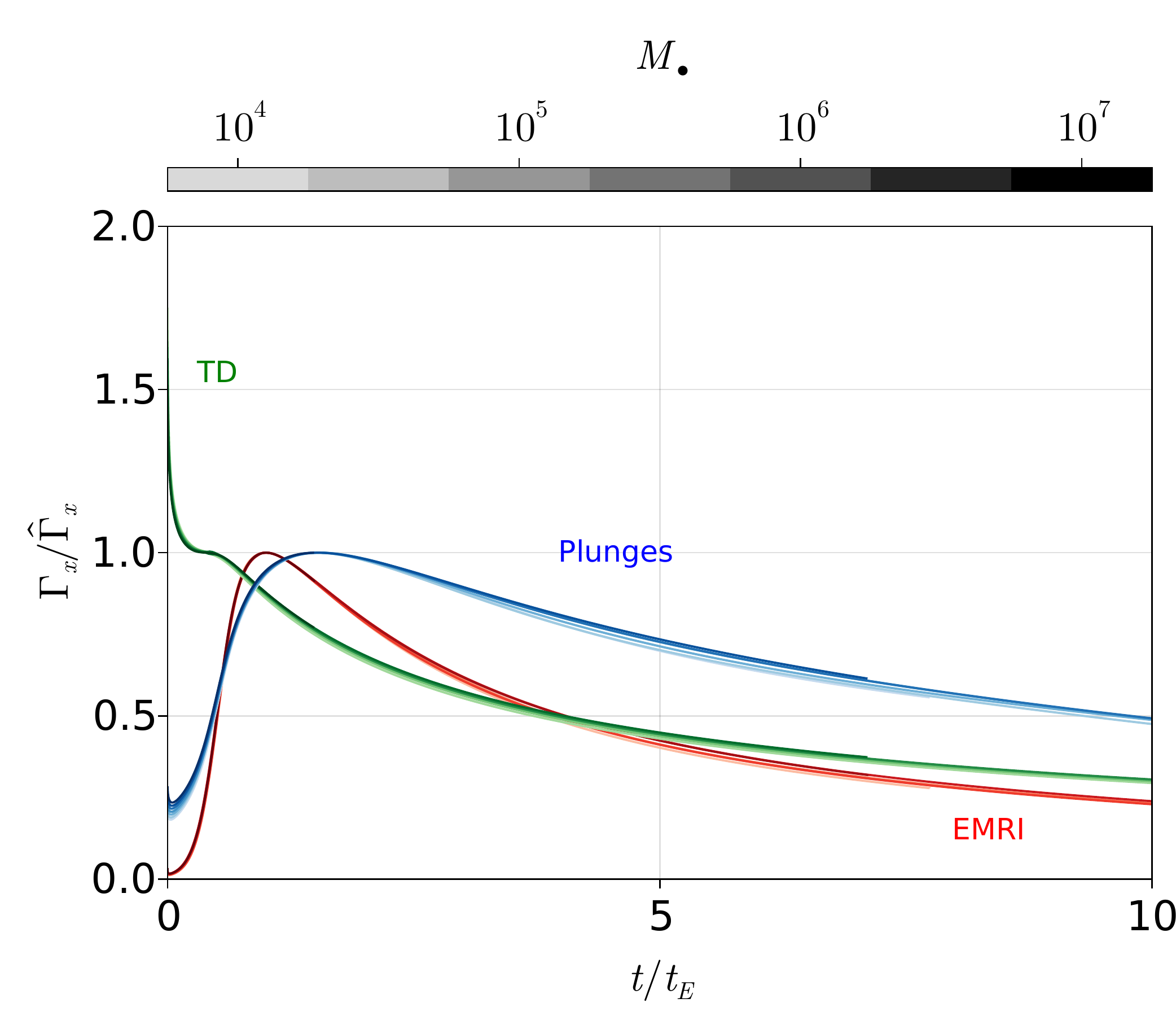}
    \caption{Rates for EMRIs (red), plunges (blue) and TD (green) formation in the simulations on the $M_\bullet - \sigma$ branch (refer to Table \ref{tbl:runs}); saturation is used to label different masses. The time evolution of the rates in all the simulations show self-similar trends once scaled properly. We express time in units of $tE$, the EMRI peak time, EMRI and Plunge rates in units of their maximum value and TDEs rate in units of their plateux value.}
    \label{fig:scaled_rates}
\end{figure}

The differential capture rate $d\Gamma/d{\rm ln}a$ can be integrated over $a$ to evaluate the formation rate of TDEs, EMRIs and plunges. As described in Section~\ref{sec:orbits}, the EMRI rate is obtain by integrating the differential rate for $a<r_\GW$, whereas the plunge rate is obtained from the integral at $a>r_\GW$. It should be noted that stars can also be swallowed by the central MBH without being disrupted. Here, however, we do refer to TDEs only, without differentiating between TDEs and swallowed stars. The fraction of the latter is simply given by the ratio of the direct capture radius over the TDE radius, i.e. $r_{\rm BH}/r_\star$. Note that as the MBH mass increases, the fraction of swallowed stars increases, accounting for all stellar captures for MBHs with $M_\bullet\gtrsim 10^8\msun$.

\subsubsection{Time evolution}
Our formulation allows us to evaluate consistently the evolution over time of the star and BH distribution function and of the rate of TDEs EMRIs and plunges,  since we do not rely on the assumption of an equilibrium solution. TDE, EMRI and plunge rates as a function of time are plotted in  Figure \ref{fig:scaled_rates} for all our simulations. For each run, the time axis has been normalised to the peak time of the EMRI rate $t_E$, whereas the peak of the EMRI and plunge rate as well as the first inflection point of the TDE rate have been normalised to unity. This specific rescaling shows that the time evolution of all rates in all simulation is almost exactly the same.
EMRIs and plunges initially grow up to a maximum value reached at slightly different times $t_P / t_E \simeq 1.8$ and then start to decay with a quasi-exponential trend. The TDE rate has no global maximum, but has a critical point around $t_{E}$. 

One can interpret this picture by considering that advection is responsible for the initial increase in the rates of the sBHs, since it moves BHs on orbits that  dominate the contribution to the flux into the loss-cone. As the distributions of the two components are rearranged, advection in the innermost regions becomes less effective until it cannot sustain the capture rate anymore, resulting in the late quasi-exponential decline. Since EMRIs are produced in the innermost region, advection slows down earlier and the EMRI peak is reached before that of plunges.
These effects have a smaller impact on the depletion rate of TDEs, which has an inflection point at an early time $t_T < t_E$. This is likely because the initial mass segregation pushes the stars out from the centre increasing the density at the influence radius, where most of the TDEs happen.

\subsubsection{Scaling with black hole mass and stellar distribution properties}
Table~\ref{tbl:runs} summarizes our findings. For each simulation we report the main parameters defining the system, namely MBH mass, $r_h$ and $\rho(r_h)$ and the main features of the TDE, EMRI and plunge rates. For each species $x$, we report the rate at the critical point (either the maximum or inflection point) $\hat \Gamma_x$, the corresponding time and the total number of events occurred during the simulation.
In order to estimate the rate of growth of the central MBH, we compute for each simulations the typical time of growth
\begin{equation}
    t_\bullet = \frac{M_\bullet}{0.5 \, m_s \, N_T + m_{\BH} \left(N_{E} + N_P\right)} \, t_{f},
\end{equation}
where $t_f$ is the total time of the simulation, which we also report in the table in units of $t_E$. The time $t_\bullet$ can be interpreted as the time the MBH needs to double its mass at the average rate of the simulations. In computing the latter, we made the simplifying assumption  that only half the mass of a star being tidally disrupted is captured by the central MBH and thus contributes to its growth. On the other hand, a sBH being captured is considered to contribute with its whole mass.

\begin{figure}
    \centering
    \includegraphics[width=0.99\linewidth]{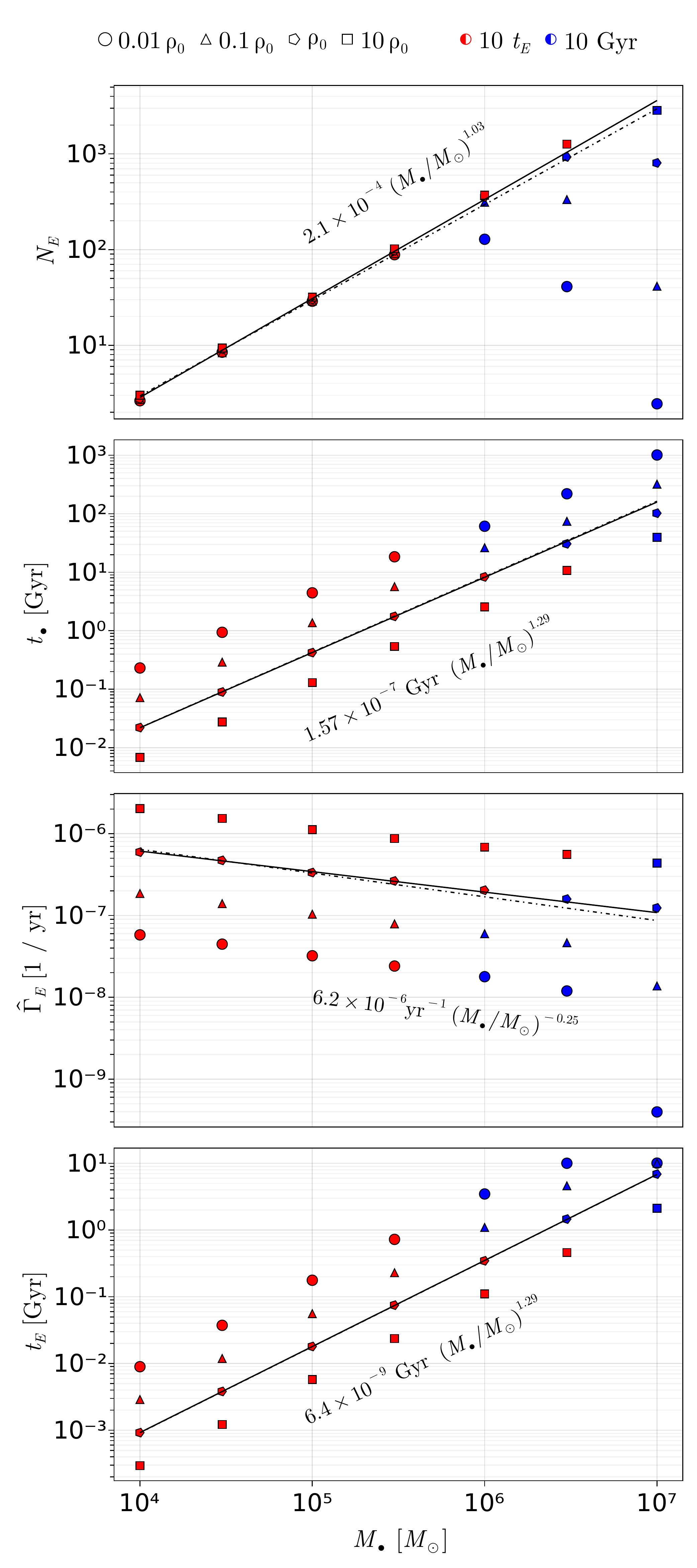}
    \caption{From top to bottom: scaling of the total number of EMRIs $N_E$, evolution timescale $t_\bullet$ of the systems, maximum EMRI rate $\hat {\Gamma}_E$ and time when it is reached as a function of $M_\bullet$ for all runs. For each plot we fit a two variables power-law in $M_\bullet$ and $\sigma$ to the $M_\bullet < 10^6 M_\odot$ simulations; we report the case $\sigma=\sigma_0$, and plot the curve in solid black. The dot-dashed line represents the best-fit powerlaw with the theoretically estimated exponents - see Table~\ref{tab:fit_M} for details. Colours mark the total time of the simulations: in red those ended at 10 times the EMRI-peak time and in blue those ended at 10 Gyr. For each mass, different shapes mark different values of the density at the influence radius $\rho(r_h)$ in units of $\rho_0$, that is the density corresponding to $\sigma_0$ from the $M_\bullet-\sigma$ relation.}
    \label{fig:scaling_with_mass}
\end{figure}

\begin{figure}
    \centering
    \includegraphics[width=\linewidth]{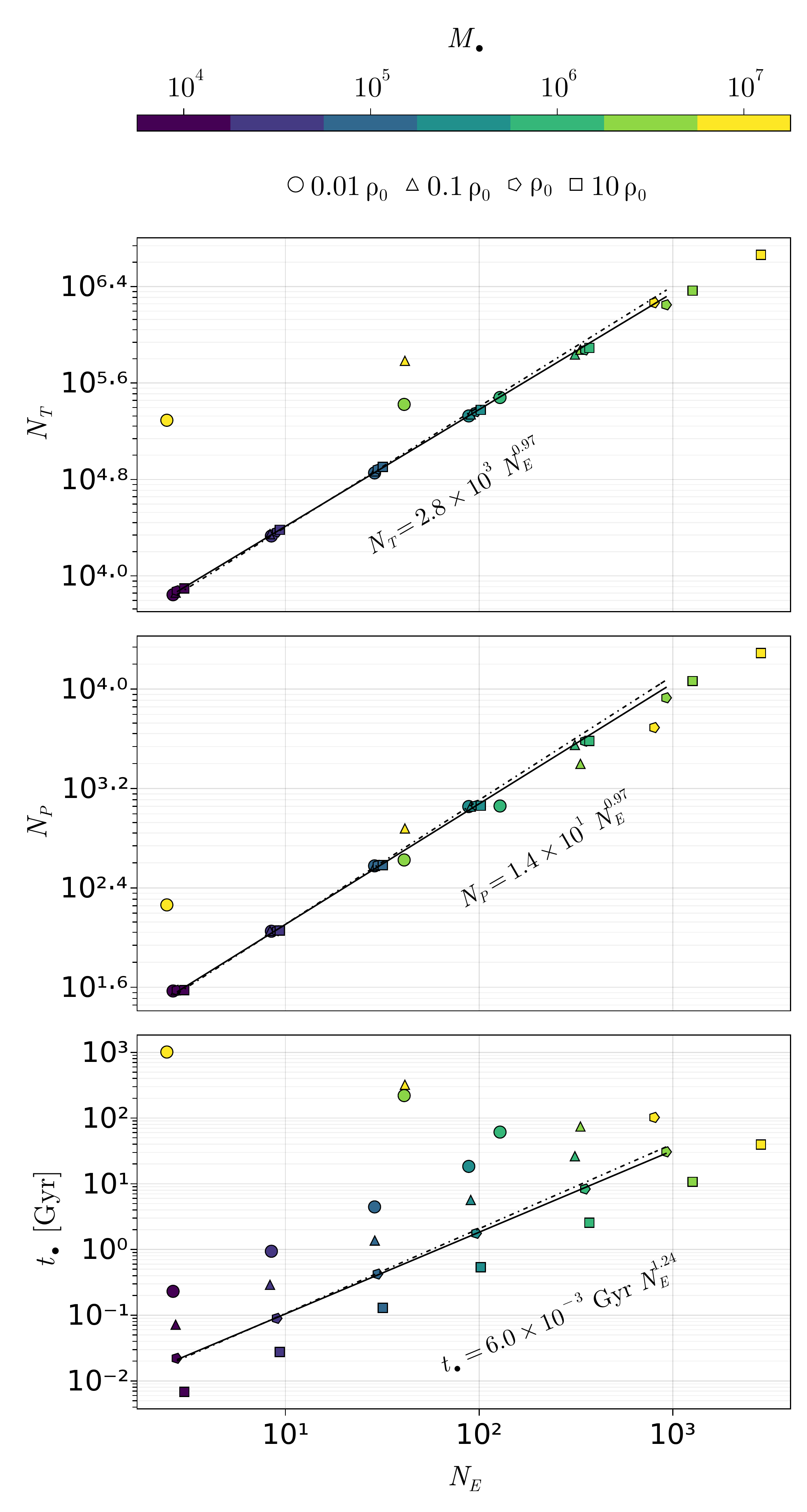}
    \caption{From top to bottom: total number of TDEs, total number of plunges and growth timescale $t_\bullet$ as a function of the total number of EMRIs for the different runs. Colours mark the central MBH mass of the simulation, according to the scale displayed by the top bar, and shapes are for different densities at a given mass (see Fig.~\ref{fig:scaling_with_mass}). For each panel, we also report the trend-line obtained by fitting a two variables power-law in $N_E$ and $\sigma$ to the $M_\bullet < 10^6 M_\odot$ simulations in solid black and the fit with the theoretically estimated trends in dot-dashed -  see Table~\ref{tab:fit_NE} for details.}
    \label{fig:scaling_with_nemri}
\end{figure}

\begin{figure}
    \centering
    \includegraphics[width=\linewidth]{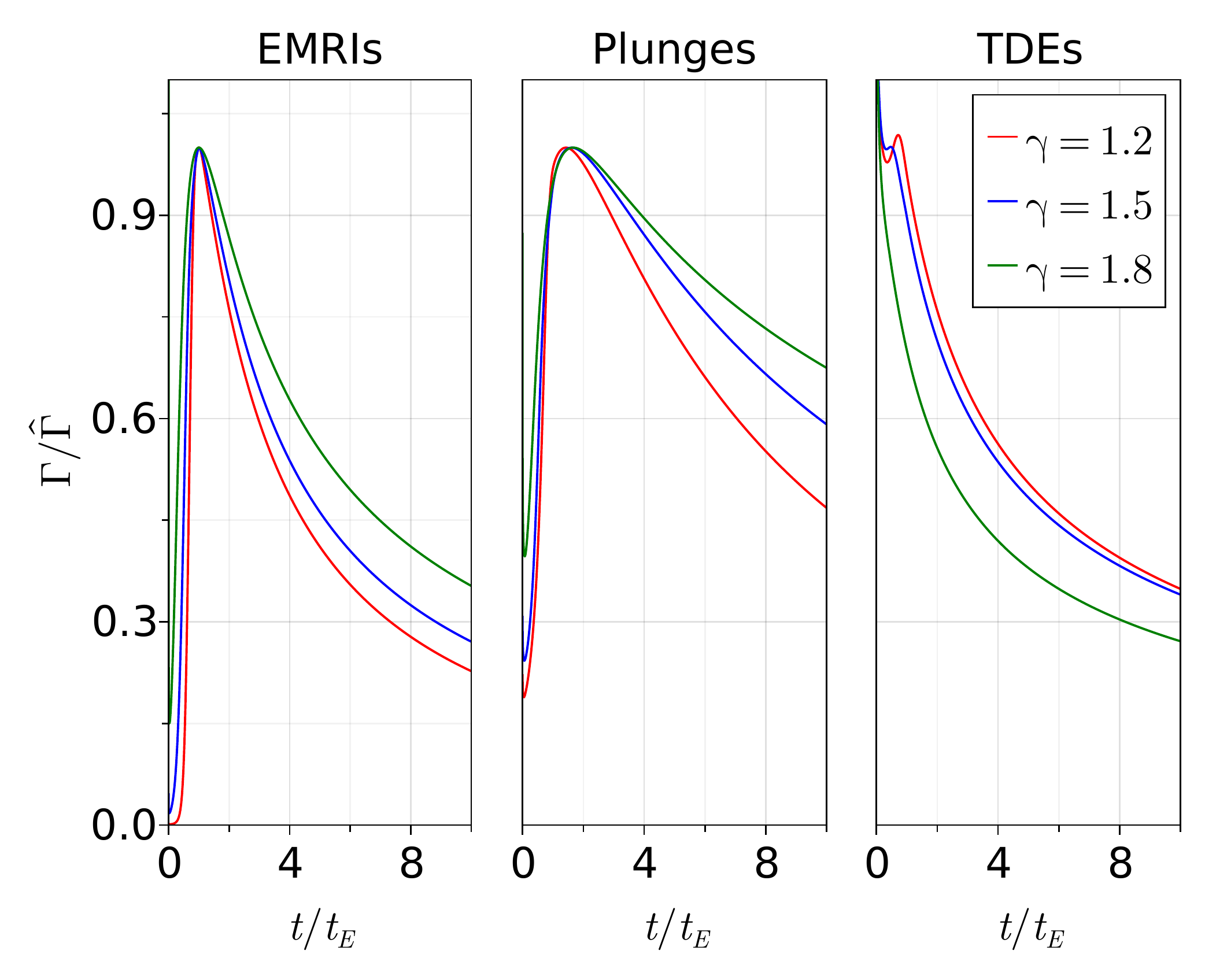}\\
    \includegraphics[width=\linewidth]{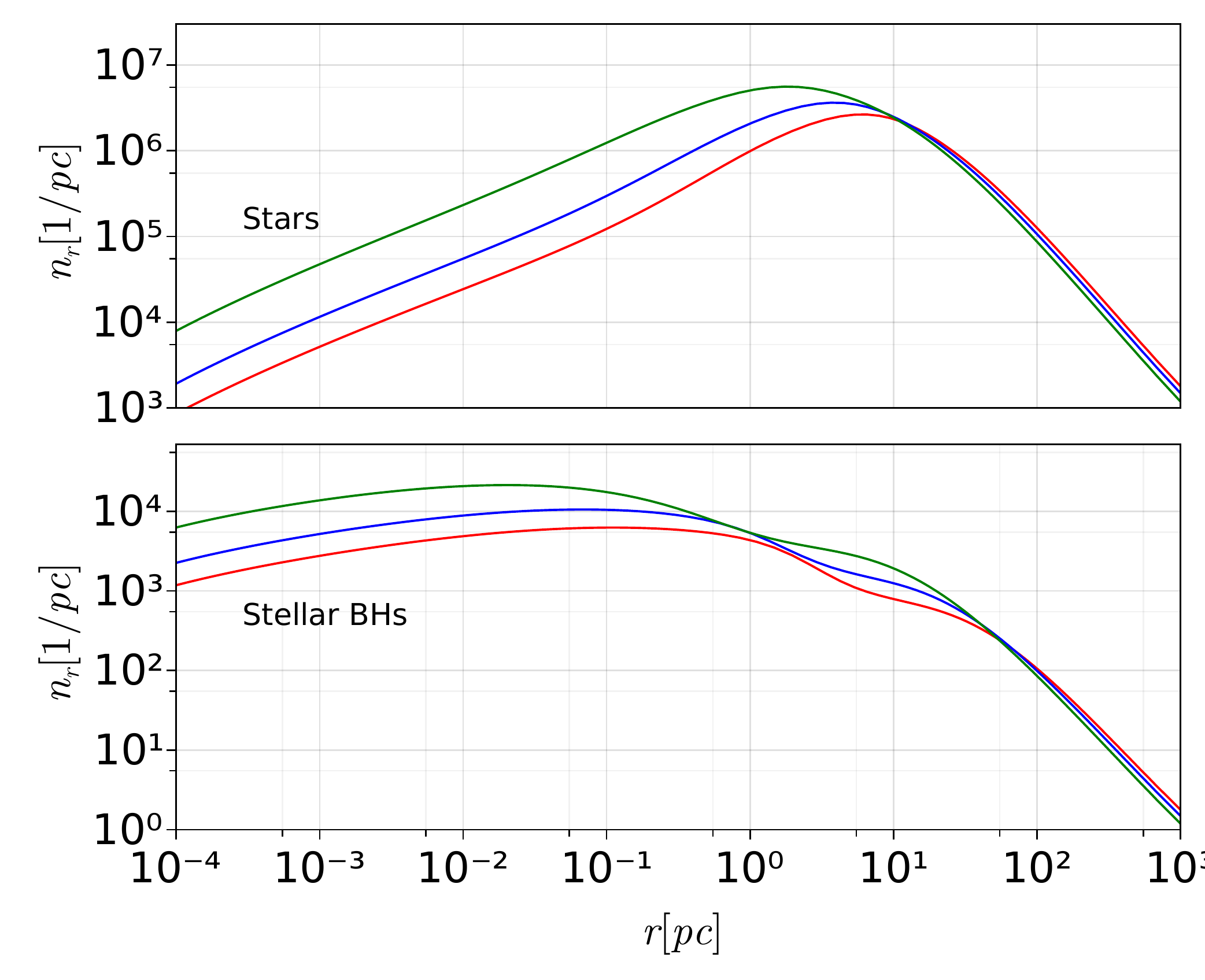}
    \caption{Results of simulations for $M_\bullet = 4\times 10^6 M_\odot$ and stellar objects initially distributed as Dehnen profiles with $\gamma = 1.2,\ 1.5,\ 1.8$ with $r_a = 4\, r_h$. The upper panel shows the rates for EMRIs, plunges and TDEs scaled as in Fig.~\ref{fig:scaled_rates}; the lower panel shows the radial number distributions at $t = 10 \, t_E$. A shallower $\gamma = 1.8$ profile corresponds to no plateau in TDEs, while $\gamma = 1.2$ has a flex. The EMRI and plunge rates show similar trends, with a slower late time decay for steep profiles.}
    \label{fig:multiple_gammas}
\end{figure}

A visual presentation of $t_\bullet$, the EMRI peak rate $\hat\Gamma_E$ and total number $N_E$ for all the simulations is given in Figure~\ref{fig:scaling_with_mass}.
It is clear that all these quantities have a power-law dependence on the central black hole mass $M_\bullet$ (and on the velocity dispersion),
which can be explained with analytical arguments once an $M_\bullet - \sigma$ relation is assumed \citep{hopman_orbital_2005}.
The average density of the system scales as
\begin{equation}
    \rho \simeq \frac{M_i}{r_a^3} \propto M_\bullet^{-2} \, \sigma^6 \propto  M_\bullet^{-0.58} \;  \left( \frac{\sigma}{\sigma_0} \right)^{6} \, ,
    \label{eq:rhoscale}
\end{equation}

which follows from the fact that the total mass of each component is proportional to $M_\bullet$ and the length scale $r_a$ is proportional to $r_h \propto M_\bullet/\sigma^2 \propto M_\bullet^{0.53}\, (\sigma / \sigma_0)^2$.
Note that in the last proportionality of equation~\eqref{eq:rhoscale} we made use of equation \eqref{eq:Msigma} to derive the $M_\bullet$ dependence. Moreover, we allow $\sigma$ to vary with respect to the scaling relation inferred value $\sigma_0$, keeping the $\sigma/\sigma_0$ dependence. This is because, for each $M_\bullet$, we explored different $\sigma$ as a way to study how the rates depend on the environment at a fixed black hole mass. Since the scale radius of the stellar distribution depends on $\sigma$, by sampling different values of $\sigma$ we can simulate environments with different typical density, as reported in the third column of Table~\ref{tbl:runs}.

Once we have the typical density, the relaxation timescale of the system at a given radius is (see \citet{merritt_dynamics_2013})
\begin{equation}
    t_{\rlx} \simeq \frac{\sigma^3}{\rho} \propto M_\bullet^{2} \; \sigma^{-3} \propto  M_\bullet^{1.29}\; \left( \frac{\sigma}{\sigma_0} \right)^{-3}.
    \label{eq:trelscale}
\end{equation}
Since the time evolution of the rate is driven by relaxation, this sets also the scaling of the quantity $t_\bullet$ plotted in the middle panel of Figure \ref{fig:scaling_with_mass}. Besides the almost perfect scaling with MBH mass, $t_\bullet$ spans a range of about 1.5dex for the different $\sigma$ adopted, in line with the scaling in equation~\eqref{eq:trelscale}.
The typical rate then behaves like
\begin{equation}
    \Gamma_{x} \propto \frac{M_{i}}{t_{\rlx}}\propto M_\bullet^{-1} \, \sigma^{3} \propto M_\bullet^{-0.29}  \;  \left( \frac{\sigma}{\sigma_0} \right)^{3} \, .
\end{equation}
This is shown for EMRIs in the lower panel of Figure~\ref{fig:scaling_with_mass}.

It can be noted that deviations from the power-law trend appear in our results. This is especially true for $N_E$ (upper panel in Figure~\ref{fig:scaling_with_mass}), which is the total number of EMRIS integrated over the simulation duration. Those deviations are due to the fact that the corresponding simulations reached the time limit $t_{f} = 10$ Gyr, possibly even before the global maximum of $\Gamma_E$ is reached.

We fit the results of our simulations to the expected power-laws in the region $M < 10^6 M_\odot$ (in order to include only simulations that run for 10 $t_E$). The general form is
\begin{equation}
    y = a_0 \; \left(\frac{M_\bullet}{M_\odot}\right)^{b_0} \; \left( \frac{\sigma}{\sigma_0} \right)^{c_0}
\end{equation}
where $a_0$ is the fitting parameter, whereas $b_0$ and $c_0$ are the power-law exponent of $M_\bullet$ and $\sigma$ predicted by the theoretical scaling for the quantity $y$ under examination. We report the trend line obtained on the $M_\bullet - \sigma$ branches in Fig. \ref{fig:scaling_with_mass}.
We also fit the general power law with three free parameters
\begin{equation}
    y = a \; \left(\frac{M_\bullet}{M_\odot}\right)^{b} \; \left( \frac{\sigma}{\sigma_0} \right)^{c}
\end{equation}
to find the best-fitting exponents and compare them to the expected theoretical scaling. The results of all fits are reported in Tab. \ref{tab:fit_M}; the fitted slopes are close to the simple scaling we derived.
\begin{table}
    \centering
    \begin{tabular}{c|cccccc}
     y & $a_0$ & $b_0$ & $c_0$ & $a$& $b$ & $c$ \\\hline
     $N_E$ & $3.0\times 10^{-4}$ & 1.00 & 0.00 & $2.0\times 10^{-3}$ & 1.03 & 0.11 \\
     $t_\bullet$ &$1.5\times 10^{-7}$ & 1.29 & -3.00 & $1.5\times 10^{-7}$ & 1.29 & -3.07\\
     $t_E$ &$6.3\times 10^{-9}$ & 1.29 & -3.00 & $6.4\times 10^{-9}$ & 1.29 & -2.97\\
     $\hat \Gamma_E$ &$9.3\times 10^{-6}$ & -0.29 & 3.00 & $6.2\times 10^{-6}$ & -0.25 & 3.09\\
    \end{tabular}
    \caption{Results of the least square fit of the data with a power law $y =~ a \,  (M_\bullet/M_\odot)^b \, (\sigma / \sigma_0)^c$. The subscript $0$ refers to the single $a_0$ parameter fit, while non dubbed quantities are the result of a three parameters fit. }
    \label{tab:fit_M}
\end{table}
\begin{table}
    \centering
    \begin{tabular}{c|cccccc}
     y & $a_0$ & $b_0$ & $c_0$ & $a$& $b$ & $c$ \\\hline
     $N_T$ & $2.5\times 10^{3}$ & 1.00 & 0.00 & $2.7\times 10^{3}$ & 0.97 & $7.1\times10^{-3}$ \\
     $N_P$ &$1.3\times 10^{1}$ & 1.00 & 0.00 & $1.4\times 10^{1}$ & 0.97 & $-8.9\times10^{-2}$\\
     $t_\bullet$ &$5.4\times 10^{-3}$ & 1.29 & -3.00 & $6.0\times 10^{-3}$ & 1.24 & -3.2\\
    \end{tabular}
    \caption{Results of the least square fit of the data with a power law $y =~ a \,  N_E^b \, (\sigma / \sigma_0)^c$. The subscript $0$ refers to the single $a_0$ parameter fit, while non dubbed quantities are the result of a three parameters fit. }
    \label{tab:fit_NE}
\end{table}

Plunges and TDEs naturally show the same scaling with $M_\bullet$ as EMRIs do since, for all species, the total number of captures is proportional to the mass of the central MBH. In Fig. \ref{fig:scaling_with_nemri}, we show the trend of $N_T$, $N_P$ and $t_\bullet$ in terms of $N_E$. The expected trends scale as
\begin{equation}
    N_T \propto N_E \quad N_P \propto N_E \quad t_\bullet \propto N_E^{1.29} \; \left( \frac{\sigma}{\sigma_0} \right)^{-3} \, .
\end{equation}
In this case we proceed as before and fit the data with a generic power-law in $N_E$ and $(\sigma/\sigma_0)$. The results of the fits are summarised in Tab. \ref{tab:fit_NE}; as in the previous case the scaling relations are close to the theoretical estimates.  

Perhaps the most important feature of these results is that $t_\bullet$, shown in Fig.~\ref{fig:scaling_with_mass}, becomes shorter than the Hubble time if $M_\bullet$ is lower than $3\times 10^5M_\odot$--$3\times10^6 M_\odot$ depending on the value of $\sigma/\sigma_0$ (i.e. depending on the initial density of the stellar distribution). This has two important implications. On the one hand, our assumption of a non evolving potential breaks down for low MBH masses, calling for a more sophisticated treatment including the time evolution of the MBH mass and, consequently, of the overall potential of the system. On the other hand, steady state EMRI rates largely used in the literature to predict LISA detections are inapplicable exactly in the mass range where LISA is most sensitive \citep[i.e. $10^5\msun<M_\bullet<10^6\msun$][]{2017PhRvD..95j3012B}, which calls for a major revision of the problem.

\subsubsection{Dependence on the slope of the stellar distribution}
In order to understand the dependence of the results on the shape of the stellar distribution, for $M_\bullet = 4 \cdot 10^6 M_\odot$ and $\sigma$ from eq.~\eqref{eq:Msigma} we performed two simulations with different initial conditions, initialising a Dehnen potential with $\gamma = 1.2$ and one with $\gamma = 1.8$. 

The main results of these runs are reported in  Table \ref{tab:various_gamam} and visualized in Fig. \ref{fig:multiple_gammas}.  In general, by increasing (in modulus) the central slope of the potential, we observe a slow-down of the late quesi-exponential decay of the three rates we consider (upper panel of Figure~\ref{fig:multiple_gammas}).
Quantitatively, considering a reference time of $10 \, t_E$ as in the previous analysis, the ratio between the total number of plunges and the total number of EMRIs decreases with $\gamma$. Since the threshold that distinguishes the two phenomena (i.e. $r_\GW$) does not depend on $\gamma$, a steeper slope implies a relatively higher number of objects inside of $r_\GW$. It is also worth noting that regardless of the initial distribution, the systems approach the same final density profiles: stars tend to $n \propto r^{-1.3}$ and stellar BHs to $n \propto r^{-1.7}$ at the centre, as shown in the lower panel of Figure~\ref{fig:multiple_gammas}.

Overall, steeper density profiles lead to higher peak rates at earlier times, resulting in an overall faster evolution of the system and mass growth of the MBH. As a direct consequence, steday state rates are even less applicable to MBHs growing in steeper density profiles.

\subsection{Comparison with literature}
The EMRI formation peak rate on the $M_\bullet-\sigma$ is best described by
\begin{equation}
    \hat \Gamma_E = 140 \ \mathrm{Gyr}^{-1} \, \left( \frac{M_\bullet}{4\cdot 10^6 M_\odot} \right)^{-0.25} \, .
\end{equation}
At the reference system $M_\bullet = 4 \cdot 10^6$, this is compatible with other estimates found in the literature \cite{preto_strong_2010,2016ApJ...820..129B,pan_formation_2021}. The best-fit power-law exponent of $\Gamma_E(M_\bullet)$ is not far from the predicted value of -0.29 (see Fig. \ref{fig:scaling_with_mass}).
The best-fit to the number of EMRIs occurring in an Hubble time, so long as $t_E\ll t_H$ (i.e. for small MBH masses, cf Figure~\ref{fig:scaling_with_mass})
is well described by
\begin{equation}
    N_E = 1.3 \cdot 10^4 \;  \left( \frac{M_\bullet}{4\cdot 10^6 M_\odot} \right)^{1.03}.
\end{equation}
\cite{pan_formation_2021} ran all the simulations for $t_f = 5$ Gyr and compared the peak rate with the average rate in the simulation
\begin{equation}
    \bar \Gamma_E = \frac{N_E}{t_f} = \frac{N_E}{5\, \mathrm{Gyr}}
\end{equation}
showing a trend compatible with ours, despite the fact of cutting the high mass end earlier and the low mass end later than in our simulations. By stopping the evolution at a time that scales with the system, the power-law describing $N_E$ emerges more clearly.

\begin{figure}
    \centering
    \includegraphics[width=\linewidth]{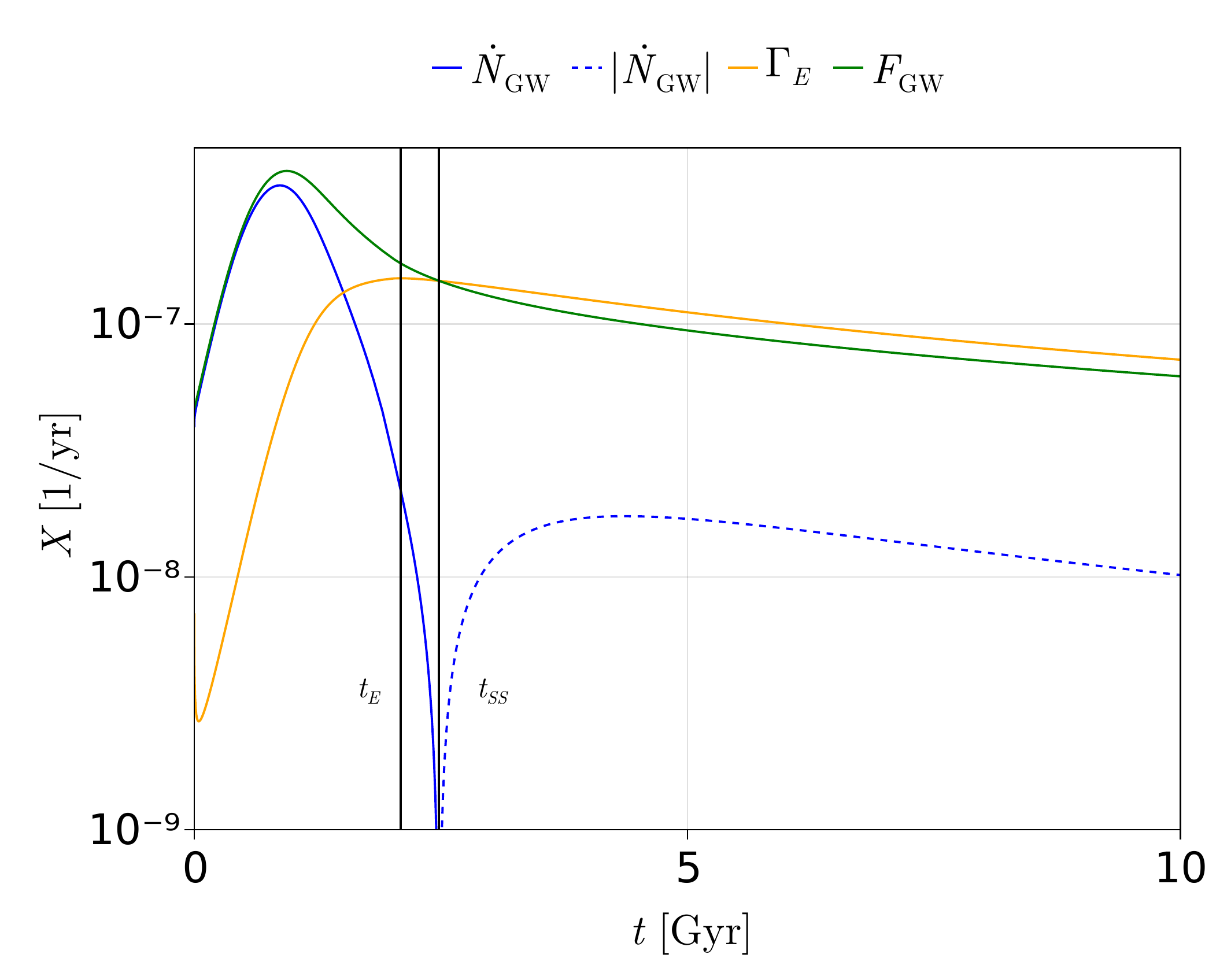}
    \caption{Inflow of particle in the $E>E_\GW$ region $\mathcal F_\GW$, total variation of sBHs in the region $\dot N_\GW$, and EMRI rate $\Gamma_E$ for the simulation with $M_\bullet = 4 \times 10^6 \, M_\odot$, $\sigma=\sigma_0$ and $\gamma = 1.5$. When the steady state approximation better describes the system $\dot N_\GW = \mathcal F_\GW - \Gamma_E = 0$ and $\Gamma_E \simeq \hat \Gamma_E$, showing why the peaks $\hat \Gamma_x$ show trends similar to steady-state rates.}
    \label{fig:GWrates}
\end{figure}

The reason why the steady-state results agree with the maximum in the steady-potential simulations resides both in the properties of the systems we consider and in the conservation of mass implied by the FP equation. Considering the region $E > E_{\GW}$  the FP equation implies that the time derivative of the number of sBHs in this region is
\begin{equation}
     \dot N_\GW = F_{\GW} - \Gamma_E
\end{equation}
where $F_{\GW}$ is the rate of particles that enter the $E>E_\GW$ region
\begin{equation}
    F_{GW} = \int_0^1 dR \; \mathcal F_E(E_\GW, R) \, .
\end{equation}
During the evolution, the number of sBHs in the region initially increases because of advection and then decreases due to gravitational captures. When it is stationary, the inflow of particles $\mathcal F_\GW$ equals the outflow of particles $\Gamma_E$, as in the steady state condition. As shown in Fig. \ref{fig:GWrates} at this time $\Gamma_E$ is around its peak.

As suggested by \cite{pan_formation_2021}, the average rate $\bar \Gamma_E$ over a suitable time interval may provide realistic formation rates without the need of an artificial cap. The capped rates derived in \cite{2017PhRvD..95j3012B} scale as $M_\bullet$ for small $M_\bullet$ and as the steady-state rates at high $M_\bullet$; this trend is qualitatively resembled by that of $\bar \Gamma_E$. While at low $M_\bullet$ the trend of $\bar \Gamma_E$ is linear, at high $M_\bullet$ it strongly depends on the curve $\Gamma_E(t)$ and is not, in general, a power-law when the Hubble time becomes comparable to $t_E$. For a better estimate, one could combine $\bar \Gamma_E$ with a probabilistic treatment of galactic encounters, since they can replenish the nuclear cluster and could possibly restart the EMRI production phase.

The time evolution of the EMRI rate may play a relevant role for LISA detection rates forecasts especially at high $M_\bullet$ - where they are strongly dependent on the initial conditions. In order to produce more reliable rates, the best strategy would be to account for realistic initial conditions and a self consistent evolution in order to directly compute the number of events produced in the evolution of a representative galactic population.



\section{Discussion and Conclusions}
\label{sec:concl}

 In this paper we developed a two-population, two-dimensional time-dependent Fokker Planck formalism to study the capture of stars and sBHs by an MBH residing at the center of a nuclear star cluster. We ran a number of simulations spanning a wide range of MBH masses and properties of the surrounding stellar and sBH distribution, keeping the gravitational potential of the system fixed. We analyzed the time dependence of the capture rate of stars and sBHs, their scaling with the properties of the systems and the growth of the central MBH. Our main findings can be summarized as follows:
\begin{itemize}
    \item Segregation dominates the evolution of the system, causing sBHs to concentrate in the center, in agreement with theoretical expectations \citep[][]{1977ApJ...216..883B}. The time dependent evolution of the distribution implies a time dependent evolution of the TDE, EMRI and plunge rates (as also shown by e.g.\citealt{2017ApJ...848...10V}). In particular EMRIs and plunges initially have a steep rise, reach a peak and then decline in a quasi-exponential fashion, whereas TDE rates experiences an initial plateau and a slow late time decay.
    \item Once normalized to the occurrence time and peak value of the EMRI rate $(t_E, \hat\Gamma_E)$, the time evolution of all species (TDEs, EMRIs, plunges) in all simulations overlap almost perfectly. We thus derived simple scalings for $t_E$ and  $\hat\Gamma_E$ which allows to reconstruct the whole time evolution of the rates of each species for any MBH mass and density of the stellar distribution.
    \item The {\it peak} of the TDE, EMRI and plunge rates is consistent with steady state estimates from the literature (e.g. \citealt{2016ApJ...820..129B}), however, those peak rates cannot be sustained indefinitely and decay over a timescale that is dependent on the MBH mass and on the properties of the nuclear star cluster (as also shown in \citealt{2022MNRAS.511.2885B} for a complete stellar mass function). 
    \item The aforementioned rates are such that the MBH doubles its mass on a timescale shorter than the Hubble time for $M_\bullet \lesssim 10^6\msun$. 
\end{itemize}

These findings have profound implications for evaluating  the rates of gravitational capture of stars and stellar mass compact objects (especially sBHs) in galactic nuclei, which are a key element for building reliable estimates of the number of EMRI events expected for future GW  missions (such as LISA) and for interpreting TDEs in electromagnetic transient surveys. Moreover, these captures can contribute significantly to the growth in mass of relatively light ($M_\bullet<10^6\msun$) MBHs, which is generally neglected in theoretical and numerical models for the evolution of MBHs along the cosmic history. 

In the literature, TDE, EMRI and plunge rates are often computed assuming steady state models for Milky-Way like systems and then scaled at lower MBH central MBH masses \citep[e.g.][]{2015ApJ...814...57M}. The underlying assumption is that a negligible fraction of the star cluster mass is captured in the loss-cone over the relevant system evolution timescale (that in this case is the Hubble time). Although this might be true for heavy MBHs (Milky Way-like or more massive), it is certainly not true for lower mass systems that evolve significantly over much shorter timescales, invalidating the steady state assumption. 

Moreover, the inner region of the nuclear star cluster is dominated by the potential of the central MBH itself, and the commonly made assumption of steady state requires that this potential remains unaltered, i.e. that the MBH mass does not grow. Our simulations indicate that for a central MBH of mass $M_\bullet \lesssim 10^6 M_\odot$ on the $M_\bullet-\sigma$ relation the mass $M_\bullet$ would change significantly over a Hubble time if all the objects entering the MBH loss-cone are accreted, invalidating \textit{a fortiori} 
the steady potential assumption. The change in mass could relevantly alter  the rates, since both the radius of influence and the capture radius depend on it. Moreover, the mass distribution of these systems at the end of the evolution has changed significantly, a fact that should be accounted by changing the critical radius for the EMRI/plunges distinction.

At present time, LISA detection rates are built under the assumption that the rate $\Gamma_E$ of a galaxy is equal to the steady-state rate computed from the mass of its central MBH \citep[][]{2017PhRvD..95j3012B} - and thus require artificial capping for small MBHs. Our findings show that those rates are likely to be biased, since the steady-state rates correspond to the maximum rate of the associated steady-potential system. More specifically, the operation of associating the rates of a steady-potential simulation to a single value of the central MBH mass is nontrivial, since they change over the same timescale. We conclude that in order to reduce the uncertainty of the detection rates it is necessary to study the whole self consistent system. As next step in the investigation of this problem, we will modify the FP formalism to allow a consistent update of the potential following the time dependent evolution of the MBH mass and of the mass distribution of the different components (stars and compact objects). 

\section*{Acknowledgements}


We thank Claudio Destri, Zehn Pan, Eugene Vasiliev and Nicholas Stone for useful discussions.
A.S. and E.B. acknowledge financial support provided under the European Union’s H2020 ERC Consolidator Grant ``Binary Massive Black Hole Astrophysics'' (B Massive, Grant Agreement: 818691).

\section*{Data Availability}
The data underlying this article will be shared on reasonable request to the corresponding author.



\bibliographystyle{mnras}
\bibliography{biblio}




\appendix
\section{Fokker-Planck coefficients}
\label{ap:FP_coeffs}
In a two components system, the diffusion coefficients that appear in equation \eqref{eq:FP} can be expressed in terms of the averaged distribution function \citep{merritt_dynamics_2013,cohn_stellar_1978}:
\begin{equation}
    \bar{f}_i(E) = \int_0^1 dR \; f(E,R)
\end{equation}
via the auxiliary functions
\begin{equation}
\begin{aligned}
  F^{0i}(E,r) &= (4\pi)^2 m_i^2 \ln\Lambda \int_{0}^E dE' \;  \bar f^i(E') \ ,\\
  F^{1i}(E,r) &= (4\pi)^2 m_i^2 \ln\Lambda \int_E^{\phi(r)} dE' \; \left(\frac{\phi-E'}{\phi-E}\right)^{1/2} \bar f^i(E') \ ,\\
  F^{2i}(E,r) &= (4\pi)^2 m_i^2 \ln\Lambda \int_E^{\phi(r)} dE'\; \left(\frac{\phi-E'}{\phi-E}\right)^{3/2} \bar f^i(E') \ .
\end{aligned}
\end{equation}
as
\begin{equation}
\begin{aligned}
  \mathcal D_{EE}^{(i)} &= \frac{8\pi^2}{3}J_c^2\int_{r_-}^{r_+} \frac{dr}{v_r} v^2\,(F^{0i}+F^{2i})
  + (i\leftrightarrow j)\ ,\\
  \mathcal D_{E}^{(i)} &= -8\pi^2 J_c^2\int_{r_-}^{r_+} \frac{dr}{v_r} F^{1i}
  +\frac{m_i}{m_j}\times (i\leftrightarrow j)\ ,\\
  \mathcal D_{ER}^{(i)} &= \frac{16\pi^2}{3}J^2\int_{r_-}^{r_+} \frac{dr}{v_r} \left(\frac{v^2}{v_c^2}-1\right)(F^{0i}+F^{2i})
  + (i\leftrightarrow j)\ ,\\
  \mathcal D_{RR}^{(i)} &= \frac{16\pi^2}{3}R\int_{r_-}^{r_+} \frac{dr}{v_r}
  \Bigg\{2\frac{r^2}{v^2} \left[v_t^2\left(\frac{v^2}{v_c^2}-1\right)^2 +v_r^2\right]F^{0i} \\
  &+ 3\frac{r^2}{v^2}v_r^2F^{1i}
  +\frac{r^2}{v^2}\left[2v_t^2\left(\frac{v^2}{v_c^2}-1\right)^2 -v_r^2 \right]F^{2i}\Bigg\}
  + (i\leftrightarrow j)\ ,\\
  \mathcal D_R^{(i)}&= -16\pi^2 R r_c^2 \int_{r_-}^{r_+}  \frac{dr}{v_r} \left(1-\frac{v_c^2}{v^2}\right) F^{1i}
  +\frac{m_i}{m_j}\times (i\leftrightarrow j)\ ,
\end{aligned}
\end{equation}
where $v_c^2 = J_c^2/r^2$ is the circular velocity at a given energy, $i$=s, $j$=BH for stars, $i$=BH, $j$=s for sBHs.

\section{Velocity dispersion of the inital conditions}
\label{ap:stellarsigma}
Assuming an isotropic distribution, in a Dehnen system with inner slope $\gamma=1.5$, the radial velocity dispersion in presence of a central MBH is \citep{tremaine_family_1994}
\begin{equation}
    \sigma^2_r = \frac{G M}{r_a} \; \left[ s_1\left( \frac{r}{r_a}\right) + \frac{M_\bullet}{M_s} \; s_2\left( \frac{r}{r_a}\right) \right]
\end{equation}
where
\begin{equation}
\begin{split}
    s_1(x) = &x^{0.5}(1+x)^{3.5} - x^{1.5}(1+x)^{2.5} \left[4\log \left(1+\frac1x\right) - \frac{10}3 \right]\\
    &-6x^{2.5}(1+x)^{1.5} + 2x^{3.5}(1+x)^{0.5} - \frac13x^{4.5}(1+x)^{-0.5}
\end{split}
\end{equation}
\begin{equation}
\begin{split}
    s_2(x) = &- \frac{256}{15} \, x^{1.5}\,(1+x)^{2.5} + \frac25 \frac{(1+x)^5}{x} - \frac83 (1+x)^4 +12 x(1+x)^3\\
    &+8x^2(1+x)^4 -\frac23x^3(1+x) \, .
\end{split}
\end{equation}
One can use this formula to compute the relation between the scale $r_a$ and the velocity dispersion of the system at different radii. Another possibility is to average this value in a central region, such as in a sphere with radius $r=r_a$
\begin{equation}
    \langle \sigma^2 \rangle_1 = \frac{\int_0^{2\,r_a} dr\, 4\pi\, r^2\, n(r) \, \sigma^2_r(r)}{\int_0^{2\,r_a} dr\,  4\pi\, r^2\, n(r)} \simeq \left (0.145 + 5.173\,\frac{M_\bullet}{M_s}\right) \frac{G\, M_s}{r_a}
\end{equation}
or within $r = 0.25\, r_a$ (which is $r_h$ in our model)
\begin{equation}
    \langle \sigma^2 \rangle_{0.25} = \frac{\int_0^{2\,r_a} dr\, 4\pi\, r^2\, n(r) \, \sigma^2_r(r)}{\int_0^{2\,r_a} dr\,  4\pi\, r^2\, n(r)} \simeq \left (0.164 + 1.729\,\frac{M_\bullet}{M_s}\right) \frac{G\, M_s}{r_a}
\end{equation}

In the systems adopted in this work we considered (neglecting the sBHs) $M_s = 20 \, M_\bullet$, so that
\begin{equation}
\langle \sigma^2_r \rangle_{0.25} \simeq 8.5 \, \frac{G \, M_\bullet}{r_a}  \qquad \langle \sigma^2_r \rangle_1 \simeq 4.6 \, \frac{G \, M_\bullet}{r_a} 
\end{equation}

\section{Further details on the algorithm}
\label{ap:Algorithm_details}
\subsection{Derivatives on the grid and time integration}
To compute the right hand side of equation \eqref{eq:FP} we opted for a flux conservative scheme, which proved to be more stable than a simple finite-difference approach. We compute each contribution to the divergence of $\mathcal{F}$ starting from the value of its components on the grid. To compute $\partial_R \mathcal F_R$ on a cell centred at ($\bar s, \bar R$), for example, we linearly interpolate the coefficients $\mathcal D^j_i$, the distribution function $f^i$ and its partial derivative $\partial_E f^i$ at the upper and lower edges of the cell $(s, R\pm\Delta R/2)$, we compute $\partial_R f^i$ at the edges with the finite difference and assemble the ingredients to give
\begin{equation}
    \partial_R \mathcal F^R_i (\bar s, \bar R) = \frac{\mathcal F^R_i(\bar s, \bar R + \Delta R/2) - \mathcal F^R_i(\bar s, \bar R - \Delta R/2)}{\Delta R} \, .
\end{equation}
An analogue scheme scheme applies to $\partial_E \mathcal F_i^R$. The boundary conditions are used to set $f^i$ and its derivatives at the corresponding locus. At the loss-cone, we rearranged the boundary condition as an equation for the partial derivative of $f^i$
\begin{equation}
    \partial_R f^i(s, \tilde R - \Delta R/2) = \frac{f^i(s, \tilde R)}{\log \tilde R/R_0} \, \frac{1}{\tilde R}
\end{equation}
where $\tilde R$ is the smallest point on the grid above the loss-cone boundary at $s$.

We integrated the equation in time using an implicit Euler scheme. The flux conservative approach we just described is formally linear in the values of the functions on the grid, that is, for each component we can write
\begin{equation}\label{eq:RHS_FP}
    \partial_E \mathcal F_E^i(s_j, R_l) + \partial_R \mathcal F_R^i(s_j, R_l) = \sum_{k, m} M_{j,l}^{i;k,m} \, f^i(s_{k}, R_{m}).
\end{equation}
This means that the expression on the left is a linear combination of the values of $f^i$. The implicit Euler scheme consists in discretising the left hand side of equation \eqref{eq:FP} at time $t$ with a forward derivative in time and plugging at the right hand side expression \eqref{eq:RHS_FP} evaluated at $t+\Delta t$. One obtains a linear system for the grid values of $f(s_j, R_l, t+\Delta t)$ that can be recast in a more familiar way by linearising the matrix $f^i(s_j, R_l, t + \Delta t)$ forming the vector $\mathfrak f^i_t$
\begin{equation}
    \mathfrak C \,  \frac{\mathfrak f^i_{t+\Delta t} - \mathfrak  f^i_t}{\Delta t} = \mathfrak M^i_t \cdot \mathfrak f^i_{t+\Delta t}
\end{equation}
where we also introduced $\mathfrak C$, the linearised version of $C(s, R)$, and $\mathfrak M$, the matrix version of the tensor $M$ in \eqref{eq:RHS_FP}. A time integration step consists in solving this equation for $\mathfrak f^i_{t + \Delta t}$, returning then to its matrix counterpart. The symbol $\cdot$ indicates matrix-to-vector multiplication.

At each step we choose the integration time step $\Delta t$ adaptively, by setting a constraint on the expected variation of $f$. In the vectorised notation
\begin{equation}\label{eq:timestep}
    \Delta t = \epsilon \, \min_{i, \textrm{grid}} \left ( \frac{\mathfrak f^t}{\dot{\mathfrak{f}}^i_t} \right ) \qquad \dot{\mathfrak f }^i_t = \frac{\mathfrak M^i_t \cdot \mathfrak f^i_t}{\mathfrak C}
\end{equation}
where $\epsilon$ is a threshold value.

\subsection{Rate across the loss-cone}
\begin{figure}
\centering
\includegraphics[width=0.7\linewidth]{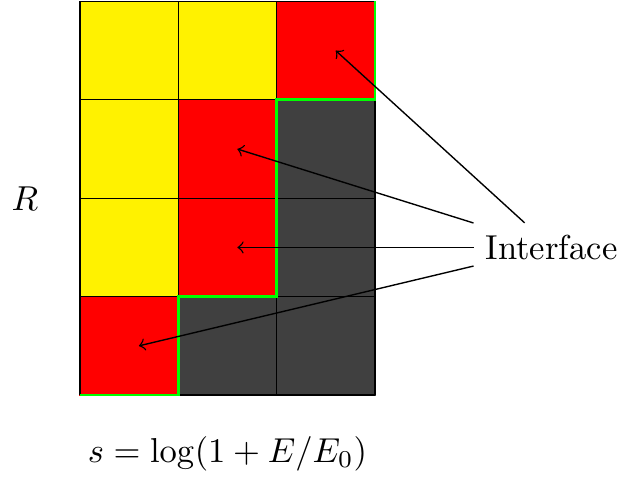}
\caption{\label{fig:Loss-cone_BND}
Discrete representation of part of the loss-cone on the simulation grid. The green line is the effective loss-cone in the simulation and the red cells are used to compute the rate of particles entering the loss-cone.}
\end{figure}

In order to compute the various rates across the loss-cone, we use the definition of the current density. Defining the curves
\begin{equation}
\begin{aligned}
    \xi_{E} : \qquad E &\to (E, R^{\BH}_{\LC}(E)) \quad &E_{\GW} &\leq E \leq E_{\BH}\\
    \xi_{P} : \qquad E &\to (E, R^{\BH}_{\LC}(E)) \quad &0 &\leq E \leq E_{GW}\\
    \xi_{T} : \qquad E &\to (E, R^s_{\LC}(E)) \quad &0 &\leq E \leq E_{s}
\end{aligned}
\end{equation}
the rates are given by
\begin{equation}
    \Gamma_{x} = \int_{\xi_{x}} d \bm {n}_x \cdot \bm{\mathcal F}
\end{equation}
where $x$ stands for the desired event and $\bm n_x$ is the normal to the curve $\xi_x$. The discretised version of this curvilinear integral can be written as
\begin{equation}
 \Gamma_x = - \sum_i \mathcal F_{R,i} \, \Delta E_i + \sum_j \mathcal F_{E,j} \, \Delta R_j
\end{equation}
where $i$ runs along the cells at the interface of the loss-cone for the $x$ event whose neighbour below is empty, $j$ along the cells whose neighbour on the right is empty (see Figure \ref{fig:Loss-cone_BND}); $\Delta E$ and $\Delta R$ indicate the sizes of the corresponding cell. 



\bsp	
\label{lastpage}
\end{document}